\documentclass[useAMS,usenatbib]{mn2e}

\usepackage{graphicx}
\usepackage{amstext}
\usepackage{amsmath}
\usepackage{natbib}
\usepackage{url}
\usepackage{setspace}
\usepackage{tabularx}
\usepackage{mathtools}

\bibpunct[ ]{(}{)}{,}{a}{}{,}     
  
\def	 \ntwoh {{\rm N$_2$H$^+$}}
\def	 \co	{{\rm C$^{18}$O}}

\def 	 \tco {{\rm $^{13}$CO}}

\def	 \ntwod {{\rm N$_2$D$^+$}}

\def	 \tone {{\rm $J=1\rightarrow0$}}
\def	 \ttwo {{\rm $J=2\rightarrow1$}}
\def	 \tthree {{\rm $J=3\rightarrow2$}}
\def	 \tonenj {{\rm $1-0$}}
\def	 \ttwonj {{\rm $2-1$}}
\def	 \tthreenj {{\rm $3-2$}}
\def 	 \kms {{\rm \,km\,s$^{-1}$}}
\def     \sol {{\rm M$_\odot$}}
\def     \arcsec {{\rm $^{\prime\prime}$}}
\def     \irdc {{\rm G035.39-00.33}}
\def     \micron{{\rm \,$\mu$m}}
\def     \ltsimm{\mathrel{\spose{\lower 3pt\hbox{$\sim$}}\raise 2.0pt\hbox{$<$}}}
\def     \gtsimm{\mathrel{\spose{\lower 3pt\hbox{$\sim$}}\raise 2.0pt\hbox{$>$}}}

\title[Complex, Quiescent Kinematics in a Highly Filamentary Infrared
  Dark Cloud]{Complex, Quiescent Kinematics in a Highly Filamentary
  Infrared Dark Cloud\thanks{Based on observations carried out with
    the IRAM 30m Telescope. IRAM is supported by INSU/CNRS (France),
    MPG (Germany) and IGN (Spain).}}  \author[Henshaw, Caselli,
  Fontani, Jim\'{e}nez-Serra, Tan, Hernandez]
      {J. D. Henshaw$^{1}$\thanks{E-mail:phy5jh@leeds.ac.uk},
        P. Caselli$^{1}$, F. Fontani$^{2}$,
        I. Jim\'{e}nez-Serra$^{3}$, J. C. Tan$^{4}$ and \newauthor
        A. K. Hernandez$^{5}$\\ $^{1}$School of Physics and Astronomy,
        University of Leeds, Leeds LS2 9JT, UK
        \\ $^{2}$INAF-Osservatorio Astrofisico di Arcetri, L.go
        E. Fermi 5, Firenze I-50125, Italy\\ $^{3}$Harvard-Smithsonian
        Center for Astrophysics, 60 Garden St., 02138 Cambridge, MA,
        USA\\ $^{4}$Department of Astronomy, University of Florida,
        Gainesville, FL 32611, USA\\ $^{5}$Department of Astronomy,
        University of Wisconsin-Madison, 475 N. Charter Street
        Madison, WI 53706-1582, USA}

\begin{document}

\date{Accepted 2012 October 24. Received 2012 October 24; in original form 2012 June 29}

\pagerange{\pageref{firstpage}--\pageref{lastpage}} \pubyear{2012}

\maketitle

\label{firstpage}

\begin{abstract}\label{abstract}
Infrared Dark Clouds (IRDCs) host the initial conditions under which
massive stars and stellar clusters form. It is therefore important to
study the kinematics, as well as the physical and chemical properties
of these regions.  Their complex structure however posits challenges
in the data interpretation. We have obtained high sensitivity and high
spectral resolution observations with the IRAM 30\,m antenna, which
allowed us to perform detailed analysis of the kinematics within one
IRDC, \irdc. This cloud has been selected for its highly filamentary
morphology and the presence of extended quiescent regions,
characteristics of dynamical youth. We focus on the \tone \ and
\tthree \ transitions of \ntwoh, \co \ (\tonenj), and make comparison
with SiO (\ttwonj) observations and extinction mapping. Three
interacting filaments of gas are found. We report large-scale velocity
coherence throughout the cloud, evidenced through small velocity
gradients and relatively narrow line widths. This suggests that the
merging of these filaments is somewhat ``gentle'', possibly regulated
by magnetic fields. This merging of filaments may be responsible for
the weak parsec-scale SiO emission detected by \citet{izaskun_2010},
via grain mantle vaporization.  A systematic velocity shift between
the \ntwoh \ (\tonenj) and \co \ (\tonenj) gas throughout the cloud of
0.18 $\pm$ 0.04 \kms \ is also found, consistent with a scenario of
collisions between filaments which is still ongoing.  The \ntwoh \
(\tonenj) is extended throughout the IRDC and it does not only trace
dense cores, as found in nearby low-mass star-forming regions. The
average H$_2$ number density across the IRDC is
$\simeq$ 5$\times$10$^4$\,cm$^{-3}$, at least one order of magnitude
larger than in nearby molecular clouds where low-mass stars are
forming. A temperature gradient perpendicular to the filament is
found. From our study, we conclude that \irdc \ (clearly seen in the
extinction map and in N$_2$H$^+$) has been formed via the collision
between two relatively quiescent filaments with average densities of
$\simeq$ 5$\times$10$^3$\,cm$^{-3}$, moving with relative velocities of
$\simeq$ 5\kms. The accumulation of material at the merging points
started $\ga$ 1\,Myr ago and it is still ongoing.
\end{abstract}


\section{Introduction}\label{introduction}

\begin{figure*}
\begin{center}
\includegraphics[trim = 50mm 50mm 20mm 10mm, clip
  ,width=1.1\textwidth]{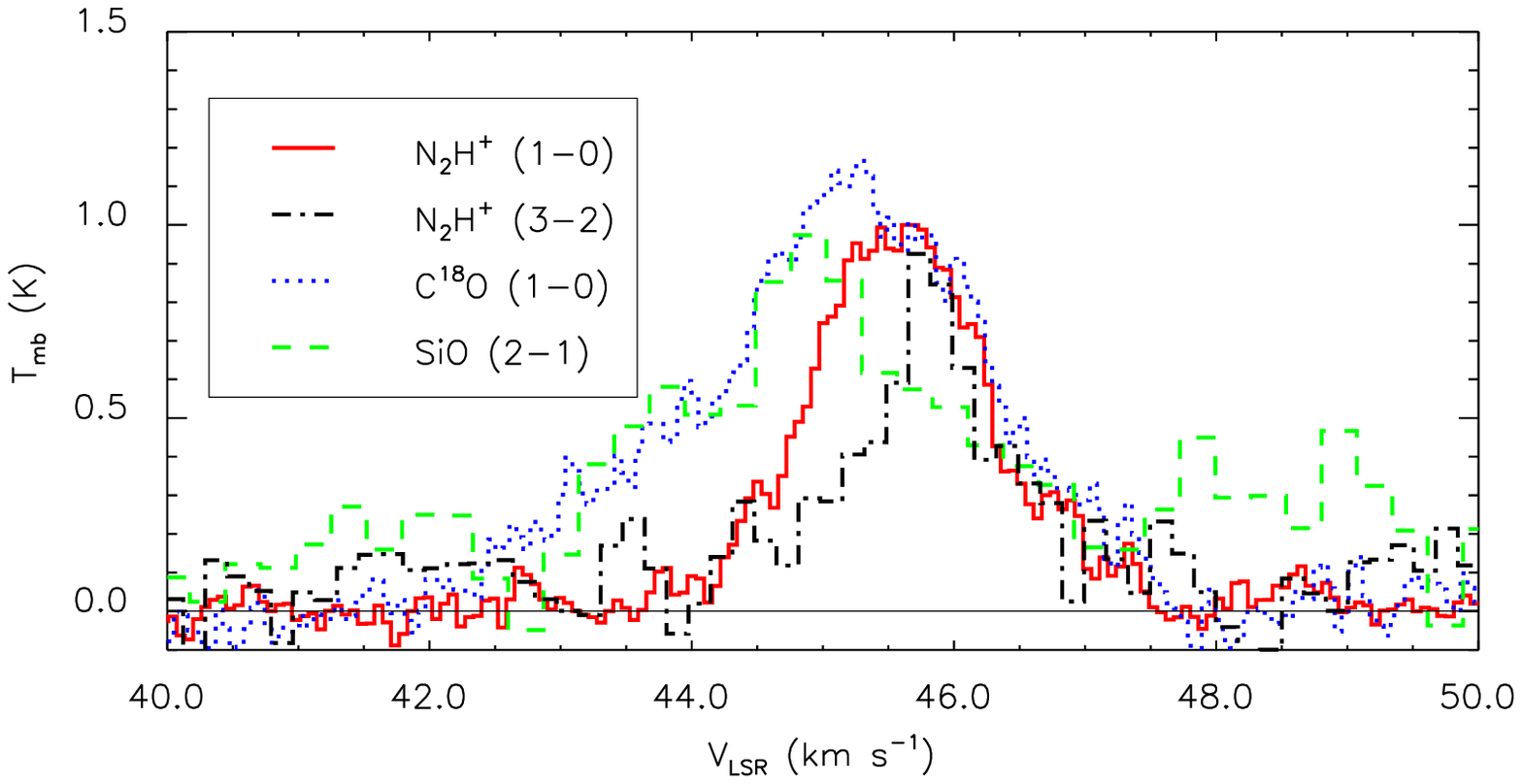}
\end{center}
\caption{Average spectra of (Red) the isolated component of \ntwoh
(\tonenj), (Black) N$_2$H$^+$(\tthreenj ), (Blue) \co \ (\tonenj), and
(Green) SiO (\ttwonj). The \ntwoh (\tonenj), (\tthreenj ) and SiO
(\ttwonj) intensities have been multiplied by factors of 2, 1.5 and
20, respectively.}
\label{spectra_plot} 
\end{figure*}
%Figure 1

In order to disentangle the complex nature of massive ($>$ 8\,\sol)
star and stellar cluster formation, we must first understand the
initial conditions under which these objects form. One way to do this
is to, firstly, observe dense clouds that have the properties
(masses, densities, temperatures etc.) required to produce massive
dense cores (hereafter MDCs). Secondly, it is preferentially required
that the MDCs are located sufficiently far away from active star
formation sites. This is to avoid complications due to, for instance,
feedback effects (such as outflows, stellar winds or UV radiation),
and to focus solely on `pristine' MDCs.

Within the past two decades, one particular group of objects has
emerged as an excellent candidate to host the initial conditions of
star and stellar cluster formation, Infrared Dark Clouds (hereafter
IRDCs). Discovered in the 1990's during mid-IR mapping of the Galaxy
with the \textit{Infrared Space Observatory} (\textit{ISO};
\citealp{perault_1996}) and the \textit{Midcourse Space Experiment}
(\textit{MSX}; \citealp{egan_1998}), IRDCs are high-extinction clouds,
seen in silhouette against the diffuse mid-IR Galactic
background. Studies have shown IRDCs are typically cold ($T<25$\,K;
e.g. \citealp{pillai_2006,peretto_2010,ragan_2011}), dense (n(H$_2$)
$\geq$ 10$^5$\,cm$^{-3}$), massive ($\sim10^2-10^5$\,M$_\odot$), have
large column densities (N(H$_2$) $\geq$ 10$^{22}$\,cm$^{-2}$)
(e.g. \citealp{carey_1998,rathborne_2006,simon_2006b,
vasyunina_2009}), and have large deuterium fractions
(e.g. \citealp{pillai_2007,chen_2010}). \citet{fontani_2011} found
large \ntwod/\ntwoh \ column density ratios towards a sample of high
mass starless cores embedded in cold IRDCs ($\sim$ 0.3 $-$ 0.7). IRDCs
have a range of morphologies, and those having a filamentary structure
are thought to be in an early stage of evolution, as described by
dynamical formation models of molecular clouds
(e.g. \citealp{van_Loo_2007,hennebelle_2008,heitsch_2009}).

The IRDC studied here, \irdc, has been selected to (i) have one of the
most extreme filamentary structures in the \citet{rathborne_2006}
sample of 38 IRDCs studied in millimetre continuum emission, (ii) have
extended quiescent regions with relatively low star formation
activity, and (iii) be relatively nearby (D = 2.9\,kpc). We have
focussed specifically on the northern portion of the cloud. The
southern portion of the cloud exhibits more tracers of star formation
activity (8\,\micron \ and 24\,\micron \ point sources), and is
thought to be in more of an advanced evolutionary stage. The portion
over which we have mapped IRDC \irdc \ contains three massive cores
from the \citet{rathborne_2006} and \citet[][\hspace{-3px}; hereafter
  BT09 and BT12 studies]{butler_2009,butler_2012}. The most massive
core (called H6 in BT09 and MM7 in Rathborne et al. 2006) has a mass
of 60\,\sol \ within a radius of 0.15\,pc (BT12) and there are no
infrared sources at the mass surface density peak. Core H5/MM6 hosts
young stellar objects and has a mass of 36\,\sol \ inside a radius of
about 0.13\,pc, while H4/MM8 is starless, with a mass of 30\,\sol
\ inside 0.2\,pc. \citet{nguyen_2011} have also mapped IRDC \irdc \ in
the far-infrared with \textit{Herschel} to investigate the star
formation activity within the cloud. They detected a total of 28
cores, 13 of which were considered to be MDCs (with masses $>$ 20
\sol, densities $>$ 2 $\times$ 10$^5$ cm$^{-3}$), potentially forming
high mass stars. This study was carried out over the full length of
the IRDC, and not just toward the northern portion selected here.

\begin{figure}
\begin{center}
\includegraphics[trim = 5mm 10mm 10mm 15mm, clip
,width=0.55\textwidth]{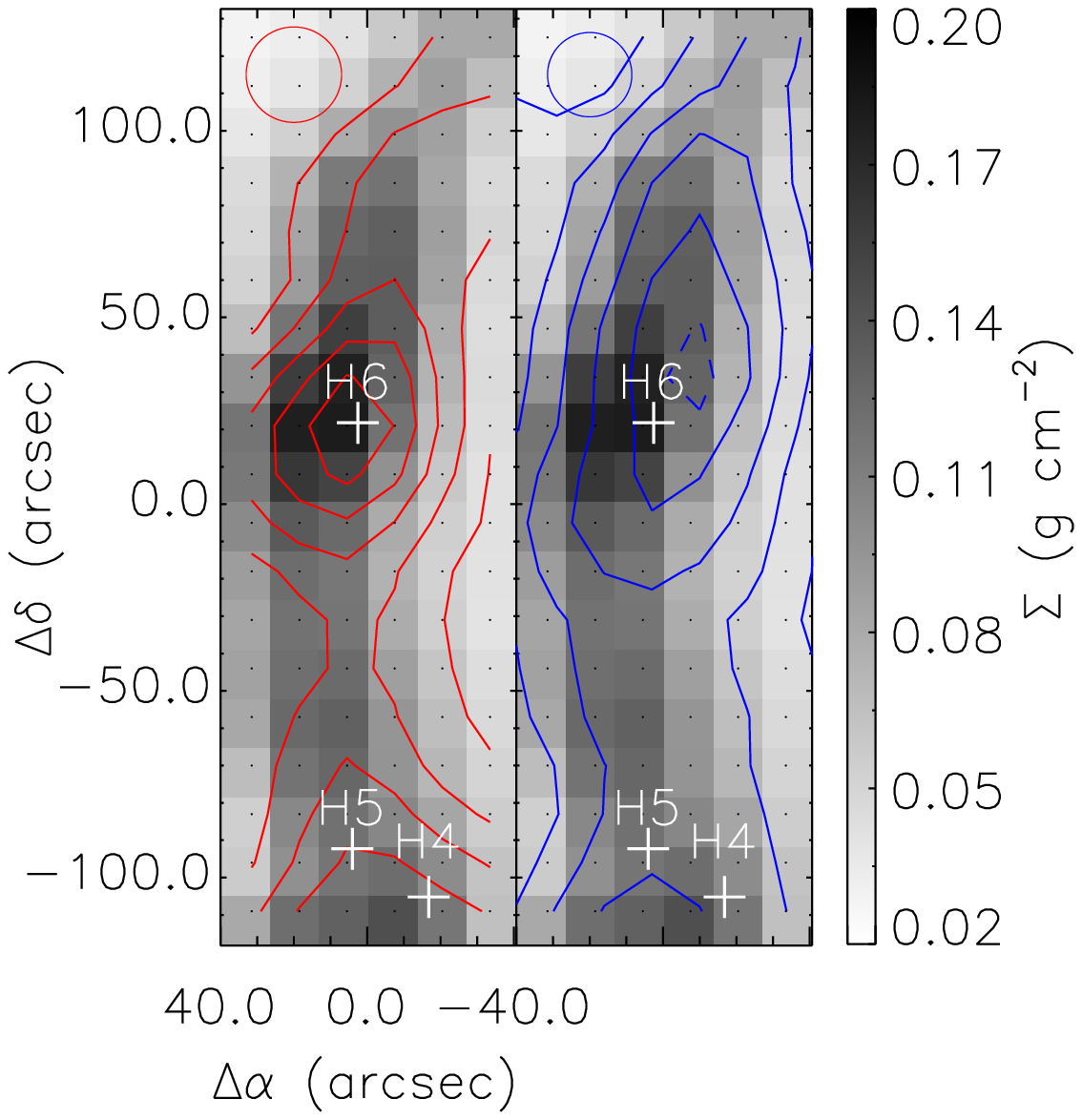}
\end{center}
\caption{Integrated intensity maps seen in contours, of (Left) \ntwoh
  \ (\tonenj) isolated component and (Right) \co \ (\tonenj), overlaid
  on the mass surface density map smoothed to a 26\arcsec
  \ beam. Contours are (Left) 10$\rm{\sigma}$ to 1.5\,K \kms, by
  5$\rm{\sigma}$, where $\rm{\sigma}$=0.05\,K \kms, (Right)
  10$\rm{\sigma}$ to 4.5\,K \kms, by 5$\rm{\sigma}$, where
  $\rm{\sigma}$=0.15\,K \kms. The dashed contour at 5.0\,K \kms \ is
  added to show the location of the peak. The red and blue circles in
  the top left of each panel refer to the half-power beam widths of
  the \ntwoh \ (\tonenj), and \co \ (\tonenj) maps (26\arcsec \ and
  23\arcsec, respectively). White crosses indicate locations of the
  cores identified by BT12 following their detection in millimetre
  continuum emission by \citet{rathborne_2006}.}
\label{int_tot} 
\end{figure}
%Figure 2

\begin{figure}
\begin{center}
\includegraphics[trim = 20mm 30mm 30mm 25mm, clip
,width=0.55\textwidth]{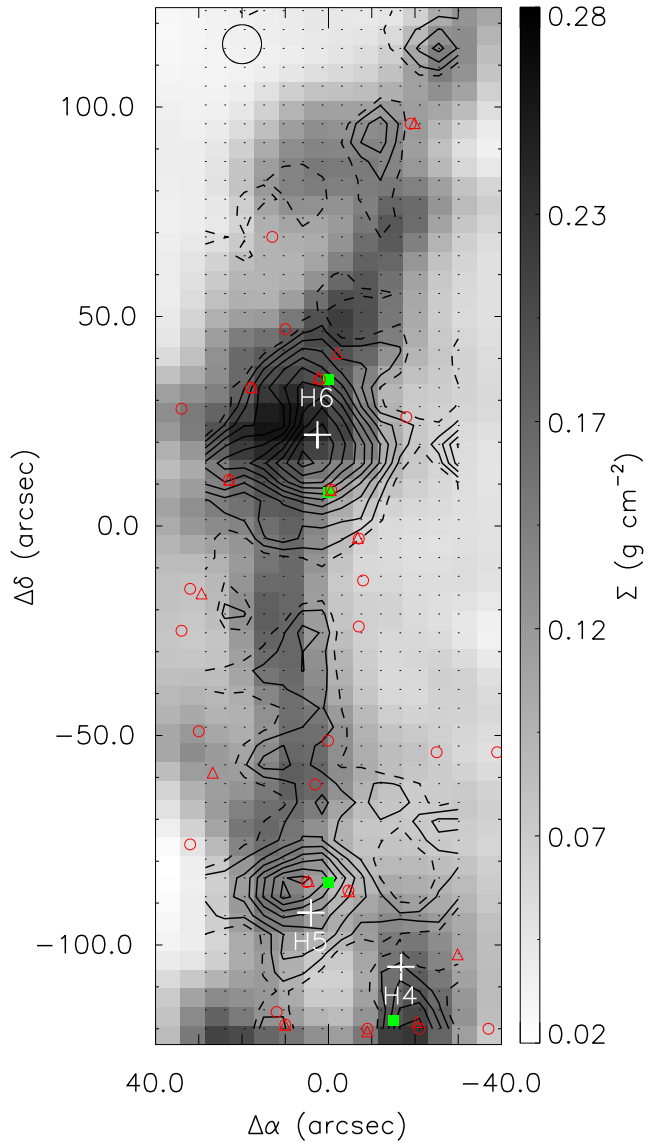}
\end{center}
\caption{Integrated intensity contours of \ntwoh \ (\tthreenj),
  overlaid on the mass surface density map smoothed to a 9\arcsec
  \ beam. Contours are 3$\sigma$ to 5.52\,K\kms, by 1$\sigma$, where
  $\sigma$=0.46K\,\kms. The dotted contour is the 2$\sigma$
  contour. White crosses indicate the positions of the massive cores
  (BT12). Colored symbols are 4.5\micron \ (green squares), 8\micron
  \ (red circles) and 24\micron \ (red triangles) sources found in the
  direction of the IRDC (see Figure\,1 of Paper I for more details on
  these sources).}
\label{int_tot_n2hp} 
\end{figure}
%Figure 3

Detailed mapping and analysis has recently been carried out towards
the northern portion of IRDC \irdc.  BT09 and BT12 used extinction
mapping in the 8\micron \ Spitzer-IRAC band on a subsample of 10 IRDCs
from the \citet{rathborne_2006} sample, providing a high angular
resolution ($\sim$ 2\arcsec) map of the mass surface density in all
cases. \citet{izaskun_2010} (hereafter, Paper I), presented the first
results from a dedicated molecular line study of this IRDC. They found
widespread SiO (\ttwonj) emission covering a large area of the
filament. This is suggestive of a large-scale shock caused by the
recent collision of molecular filaments, which may have formed the
IRDC. \citet{hernandez_2011} (hereafter, Paper II) measured
significant levels of CO freeze-out across the IRDC, indicative of
cold dense gas.  \citet{hernandez_2011a} studied the dynamics of the
IRDC using \tco \ (\tonenj) from the Galactic Ring Survey (Jackson et
al. 2006). Their results suggested the possible presence of
dynamically important surface pressures around the cloud, perhaps
indicating it was relatively young and had not yet settled into virial
equilibrium. However, \citet{hernandez_2012a} (hereafter, Paper III),
revisited this question of virial equilibrium, now using higher
resolution \co \ data and the improved extinction map of BT12. In
contrast to \citet{hernandez_2011a}, they found the IRDC exhibited
kinematics consistent with models of filamentary virial equilibrium,
which requires the cloud to be older than one dynamical time,
$\sim$1\,Myr.

In this paper (Paper IV of this series), we focus mainly on \ntwoh
\ (\tonenj), \ntwoh \ (\tthreenj ) and \co \ (\tonenj). \ntwoh \ is a
known tracer of cold dense regions ($>$ 10$^4$\,cm$^{-3}$), and in
low-mass star-forming regions it therefore typically traces the dense
cores (e.g. \citealp{caselli_2002b}). \co \ is a tracer of more
diffuse material ($\geq$ 10$^3$\,cm$^{-3}$). Due to the fact that it
is readily abundant in molecular clouds, \co \ gives more of an
indication of the general morphology of the whole cloud. Unlike
$^{12}$CO and $^{13}$CO, the relatively less abundant \co \ is
typically optically thin in molecular clouds, making it an ideal
diagnostic tool to study the kinematics of the lower density
material. In addition to these tracers we also include the mass
surface density derived by BT12 as a tracer of the dense material in
the filament. The SiO (\ttwonj) emission, studied extensively in Paper
I, is also used to compare our kinematic results, spatially, with the
shocked gas within the filament. Finally, the \tthree \ transition of
\ntwoh \ is used together with the \tone \ transition to estimate the
number density across the filament.

Details of the observations can be found in
Section\,\ref{observations}, with the results (integrated intensity
maps, channel maps, kinematics, column and number density) in
Section\,\ref{results}. A discussion of our results can be found in
Section\,\ref{discussion} and conclusions are in
Section\,\ref{conclusions}.  Appendix\,\ref{GGF} outlines the fitting
method used to extract information from this data, whereas Appendix B
reports tables with the Gaussian and hyperfine structure fits in
selected positions across the filament.

\section{Observations}\label{observations}

\begin{figure*}
\begin{center}
\includegraphics[trim = 0mm 0mm 10mm 10mm, clip
, width=1.0\textwidth]{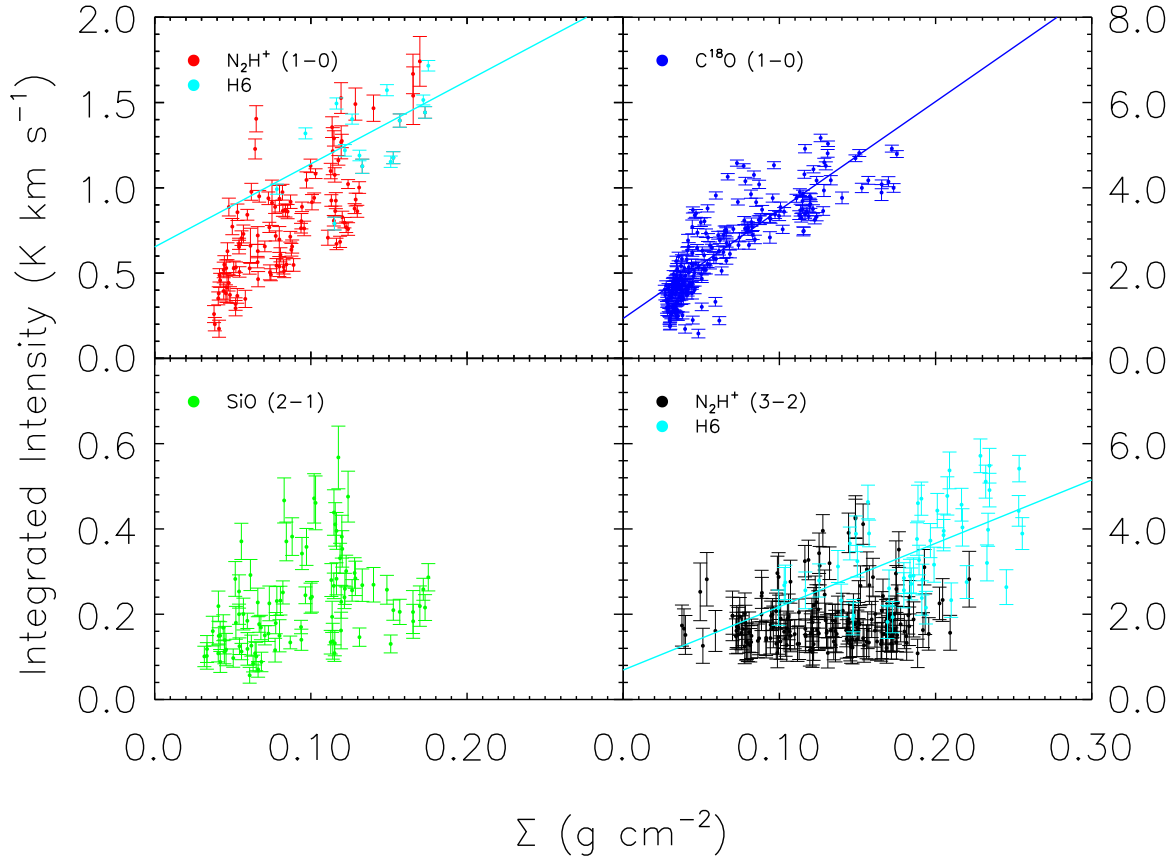}
\end{center}
\caption{Integrated intensity as a function of mass surface density
  for the (Red) \ntwoh \ (\tonenj) isolated component, (Blue) \co
  \ (\tonenj), (Green) SiO (\ttwonj) (all smooted to 26\arcsec), and
  the 9\arcsec \ resolution (Black) \ntwoh \ (\tthreenj). The \ntwoh
  \ integrated intensity spectra within the starless core H6 are
  displayed in Cyan.}
\label{ii_scatter}
\end{figure*}
%figure 4

The \ntwoh \ (\tonenj), \ntwoh \ (\tthreenj), \co \ (\tonenj), and SiO
(\ttwonj) observations were mapped with the Instituto de
Radioastronom\'{i}a Milim\'etrica (IRAM) 30-m telescope at Pico
Veleta, Spain. The \ntwoh \ observations were carried out in August
2008 and February 2009, \co \ (\tonenj) in August 2008, and SiO
(\ttwonj) observations in December 2008 and February 2009. The
large-scale images were obtained with the On-The-Fly (OTF) mapping
mode using the offsets (1830$^{''}$, 658$^{''}$) for the 2008
observations, and (300$^{''}$, 0$^{''}$) for observations carried out
in 2009 (i.e. for part of the \ntwoh \ and SiO data, whose emission is
concentrated in the IRDC), as off-positions. The central coordinates
of the maps are $\alpha$(J2000) = 18$^h$57$^m$08$^s$, $\delta$(J2000)
= 02$^{\circ}$10$'$30${''}$ (\textit{l} = 35$^{\circ}$.517, \textit{b}
= -0$^{\circ}$.274). All observations were carried out using the old
ABCD receivers.  The VErsatile SPectrometer Assembly (VESPA) provided
spectral resolutions between 20 and 80\,kHz. Information on the beam
sizes, frequencies and velocity resolutions can all be found in
Table\,\ref{obs_table}. Typical system temperatures range from 100-200
K for all line observations except for \ntwoh \ (\tthreenj) which has
a value of T$_{\rm{SYS}}$ $\sim$ 750\,K. All intensities were
calibrated in units of antenna temperature, T$^{*}_{\rm{A}}$. To
convert these intensities into units of main-beam brightness
temperature (assuming a unity filling factor), T$_{\rm{MB}}$, we have
used the beam and forward efficiencies listed in Table\,\ref
{obs_table}.  Saturn was used to calculate the focus, and pointing was
checked every 2\,hours on G34.3+0.2. The data were calibrated with the
chopper-wheel technique \citep{kutner_1981}, with a calibration
uncertainty of $\sim$20\%.

As in Paper III, we utilize the 8\,\micron \ extinction derived mass
surface density map of BT12, as modified by \citet{jouni_2012}
(hereafter, KT12) to include corrections for the presence of the NIR
extinction derived IRDC envelope. The mass surface density map has 2"
resolution, which we have smoothed to 26" (unless otherwise stated),
so as the data is of comparable resolution to the \ntwoh \ (\tonenj)
and \co \ (\tonenj) observations.

\begin{table*}
\caption{Observed molecular transitions, line frequencies, spectral
  resolutions, telescope beam sizes, beam correction factors for the
  ABCD receivers at the IRAM 30\,m telescope.\vspace{0.6cm}}
\centering \Large
%\begin{minipage}{\textwidth}
%\begin{tabularx}\textwidth}
\begin{tabular}{c c c c c c}
\hline\hline
Transition & Frequency & Spectral                 & Beam Size & Beam         & Forward  \\ [0.5ex]
                   & (MHz)         & Resolution (\kms) & (arcsec)      & Efficiency  & Efficiency \\ [0.5ex]
\hline
SiO \ttwo & 86846.9600$^a$ & 2.68$\times$10$^{-1}$ & 28 & 0.77 & 0.98 \\
\ntwoh \tone & 93173.7637$^b$ & 6.28$\times$10$^{-2}$ & 26 & 0.76 & 0.98\\
\co \tone & 109782.1730$^c$ & 5.33$\times$10$^{-2}$ & 23 & 0.73 & 0.97 \\
\ntwoh \tthree & 279511.8320$^b$ & 4.19$\times$10$^{-2}$ & 9 & 0.49 & 0.90 \\ [1ex]
\hline\hline
\end{tabular}
%\end{tabularx}

\begin{minipage}{\textwidth}

\small{$a$: \citet{izaskun_2010},}\\
\small{$b$: \citet{pagani_2009},}\\
\small{$c$: \citet{cazzoli_2003}}

\end{minipage}
\label{obs_table}
\end{table*}

\section{Results}\label{results}

\subsection{Spectra and integrated intensities}\label{spectra}

The spectra of the isolated \ntwoh \ (\tonenj) hyperfine component
(F$_1$, F = 0,1 $\rightarrow$ 1,2), the \ntwoh \ (\tthreenj ), the \co
\ (\tonenj), and the SiO (\ttwonj), averaged across the whole mapped
area, are shown in Figure\,\ref {spectra_plot}. The spectra are
plotted in units of main beam brightness temperature, with the \ntwoh
\ (\tonenj), \ntwoh \ (\tthreenj ) and SiO (\ttwonj) being multipled
by factors of 2, 1.5 and 20, respectively, for comparison.  The four
molecular gas tracers have different centroid velocities, with the
highest density tracer, \ntwoh \ (\tthreenj ), peaking at the largest
velocity, followed by \ntwoh \ (\tonenj), \co \ (\tonenj) and SiO
(\ttwonj). The SiO (\ttwonj) shows the broadest width, with emission
present up to velocities of about 50\kms. The \co \ line encloses the
emission of the other species (except for the high velocity emission
of the SiO (\ttwonj)) and presents an asymmetric profile indicative of
multiple velocity components along the line of sight. Although shifts
in velocities relative to the SiO line are expected given that SiO
typically traces shocked material (e.g. \citealp{izaskun_2009}), {\it
  the significant shifts in centroid velocities observed in the more
  quiescent molecular cloud tracers \co \ and \ntwoh \ represent a
  clear difference from low-mass star-forming regions}
\citep{walsh_2004,hacar_2011}.

Figure\,\ref{int_tot} presents integrated intensity maps of the \tone
\ transitions of both \ntwoh \ (Red contours; F$_1$, F = 0,1
$\rightarrow$ 1,2 hyperfine component), and \co \ (Blue contours),
convolved with the respective beams (see Table\,\ref {obs_table}) and
using a pixel size of half the respective beam sizes. In each map, the
spectra have been integrated between 40 and 50\,\kms, to cover all the
emission (see Figure\,\ref{spectra_plot}).  The \ntwoh \ (\tonenj) and
\co \ (\tonenj) maps have been superimposed on the mass surface
density map smoothed to 26\arcsec \ angular resolution. Overlaid are
white crosses to indicate the location of the three massive cores seen
in the millimetre continuum emission \citep{rathborne_2006} and in the
unsmoothed extinction map \citep{butler_2009}.  The integrated
intensity map of the \ntwoh \ (\tthreenj) emission is shown in
Figure\,\ref {int_tot_n2hp}, overlaid on the mass surface density map
smoothed to the IRAM 9\arcsec \ beam at the frequency of this
transition.

From Figures\,\ref {int_tot} and \ref {int_tot_n2hp} we find that {\it
  the \ntwoh \ {\rm (\tonenj)} and {\rm (\tthreenj)} emitting regions
  are extended across a large portion of the filament}. This is
different from low-mass star-forming regions, where \ntwoh \ mainly
traces the dense cores \citep{caselli_2002b, andre_2007,
  friesen_2010}, whereas filaments can be seen with lower density
tracers such as \co \ (\tonenj) \citep{mizuno_1995, hacar_2011}. Given
that the critical density of the \ntwoh \ (\tonenj) transition is
$\simeq$\,10$^5$\,cm$^{-3}$, this is a clear indication that the
overall number density of this IRDC is similar to the average density
of nearby low-mass dense {\it{cores}} (i.e. at least one order of
magnitude higher than the average density of the clouds within which
nearby low-mass cores are embedded, $\leq$10$^3$\,cm$^{-3}$;
e.g. Pineda et al. 2008). \ntwoh \ (\tthreenj) peaks towards two out
of the three massive cores mentioned above (white crosses in
Figure\,\ref{int_tot_n2hp}). The peak in both transitions of \ntwoh
\ is seen to be coincident with the position of core H6, whereas the
\co \ (\tonenj) peaks away from this (within $\sim$ 1 beam size). This
is due to the freeze-out of CO, most prominent towards core H6 (Paper
II). Emission from both \ntwoh \ transitions traces the morphology of
the extinction fairly well, whereas the emission from the \co
\ (\tonenj) covers a larger area.

Figure\,\ref {ii_scatter} shows the total integrated intensities of
the \tone \ (F$_1$, F = 0,1 $\rightarrow$ 1,2) and \tthree
\ transitions of \ntwoh, \co \ (\tonenj), and SiO (\ttwonj) as a
function of mass surface density. Intensities are integrated over a
velocity range of 40-50\,\kms. To highlight the properties of the most
massive core in the region, \ntwoh \ data from core H6 have different
colors\footnote{In order to separate this core from the rest of the
  data we have taken the 5$\sigma$ level of the \ntwoh \ (\tonenj) and
  \ (\tthreenj) intensity (between -20.0 $\leq$ $\Delta\alpha$ $\leq$
  20.0 and -10.0 $\leq$ $\Delta\delta$ $\leq$ 50.0), and attribute
  this to the core.}. Significant correlation (regression value,
r-value = 0.8) is found between \ntwoh \ (\tonenj) integrated
intensity and mass surface density, over the full cloud (including
core H6). This is to be expected, as the \ntwoh \ is tracing the dense
gas. A significant correlation is also found across core H6 (r-value =
0.6), despite the fewer data points. For the \ntwoh \ (\tthreenj) we
have compared the line integrated intensity and mass surface density,
both smoothed to 9\arcsec . The correlation for \ntwoh \ (\tthreenj)
is however not as strong (r-value = 0.5), over the full cloud,
probably due to the fact that this line is not just sensitive to
number density but also to temperature (see
e.g. \citealp{pagani_2007}), and warm gas may not be correlated with
the mass surface and number density (see Section \ref{vol den}).  A
slightly stronger correlation is present toward core H6 (r-value =
0.55), where the line is brightest. There is also a strong correlation
(r-value = 0.8) between the \co \ (\tonenj) integrated intensity and
the mass surface density. A weaker correlation (r-value = 0.45)
between the SiO (\ttwonj) integrated intensity and the mass surface
density is observed. This, along with the fact that the centroid
velocities of the SiO and \ntwoh \ transitions are not coincident (see
Figure\,\ref {spectra_plot}), suggests that the narrow SiO (\ttwonj)
emission is mainly associated with the lower density gas traced by \co
\ (n$_H$ $\sim$ 10$^{3}$\,cm$^{-3}$), despite the relatively high
critical density of the transition (n$_{cr}$ $\sim$ 10$^5$
cm$^{-3}$). Therefore, {\it we expect the narrow SiO {\rm (\ttwonj)}
  lines to be subthermally excited, which would explain their low
  brightness} (Paper I).

\subsection{Kinematics}\label{kinematics}

\subsubsection{Multiple velocity components}\label{multiple vel}

\begin{figure*}
\begin{center}
\includegraphics[trim = 0mm 0mm 65mm 30mm, clip ,height =
  0.9\textheight,width=0.75\textwidth]{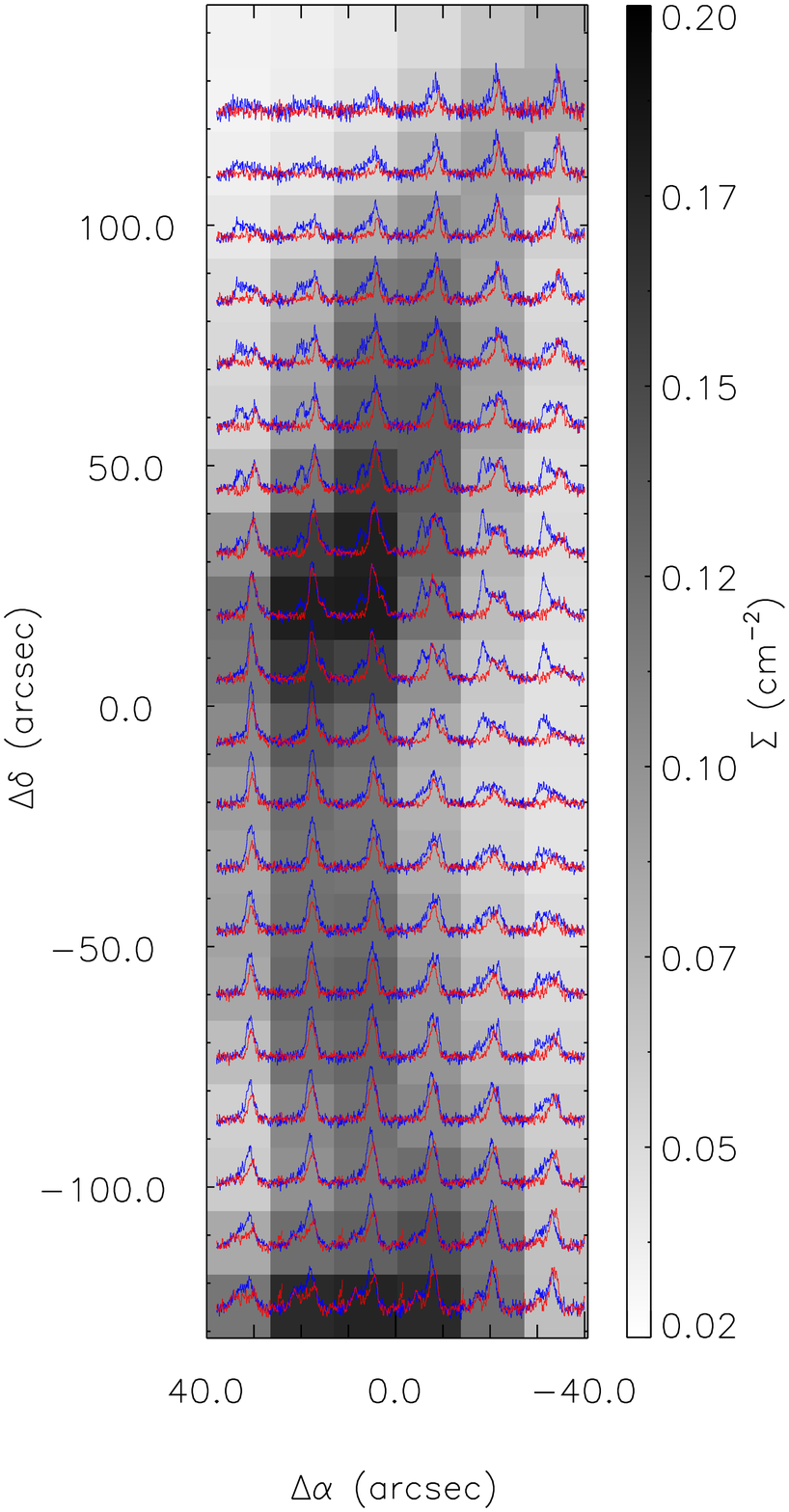}
\end{center}
\caption{Spectra of the (Red) \ntwoh \ (\tonenj) (F$_1$, F = 0,1
  $\rightarrow$ 1,2) and (Blue) \co \ (\tonenj) at all positions in
  the cloud, overlaid on the mass surface density map of KT12. Spectra
  are shown between 41-49\,\kms \ and -0.1--1.2\,K. The intensity of
  the \ntwoh \ (\tonenj) has been multiplied by a factor of 2 for
  clarity. The \co \ (\tonenj) has been smoothed to 26\arcsec, and
  regridded this to 13\arcsec, to make direct comparison with the
  \ntwoh \ (\tonenj) data.}
\label{spectra_map}
\end{figure*}
%figure 5

\begin{figure*}
\begin{center}
\includegraphics[trim = 0mm 10mm 0mm 35mm, clip
  ,width=1.\textwidth]{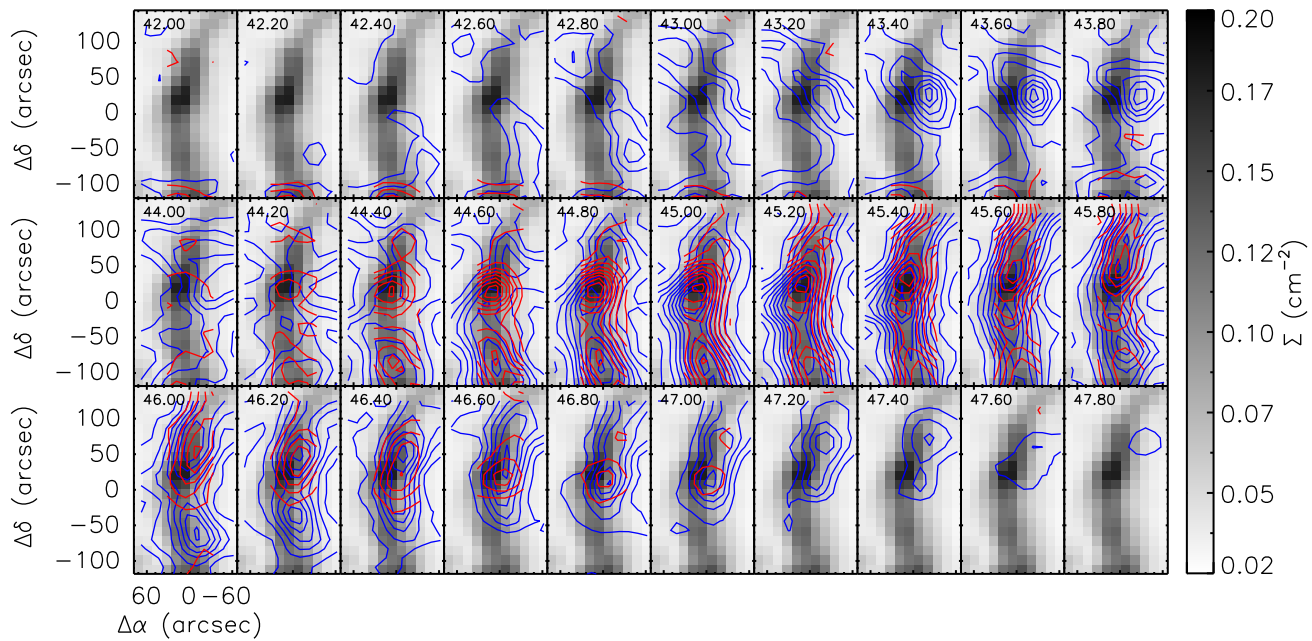}
\end{center}
\caption{G035.39-00.33 channel maps of (Red) \ntwoh \ (\tonenj)
  (F$_1$, F = 0,1 $\rightarrow$ 1,2) and (Blue) \co \ (\tonenj),
  overlaid on the mass surface density map. Intensity maps integrated
  from 43\kms \ to 48\kms \ and in steps of 0.2\kms. Contours for
  \ntwoh \ (\tonenj) in all cases are from 3$\sigma$, increasing in
  steps of 3$\sigma$ (0.02 K\kms). Contours for \co \ (\tonenj) in all
  cases are from 3$\sigma$, increasing in steps of 3$\sigma$ (0.04
  K\kms).}
\label{channel_maps}
\end{figure*}
%figure 6

\begin{figure*}
\begin{center}
\includegraphics[trim = 0mm 5mm 0mm 0mm, clip
  ,width=0.8\textwidth]{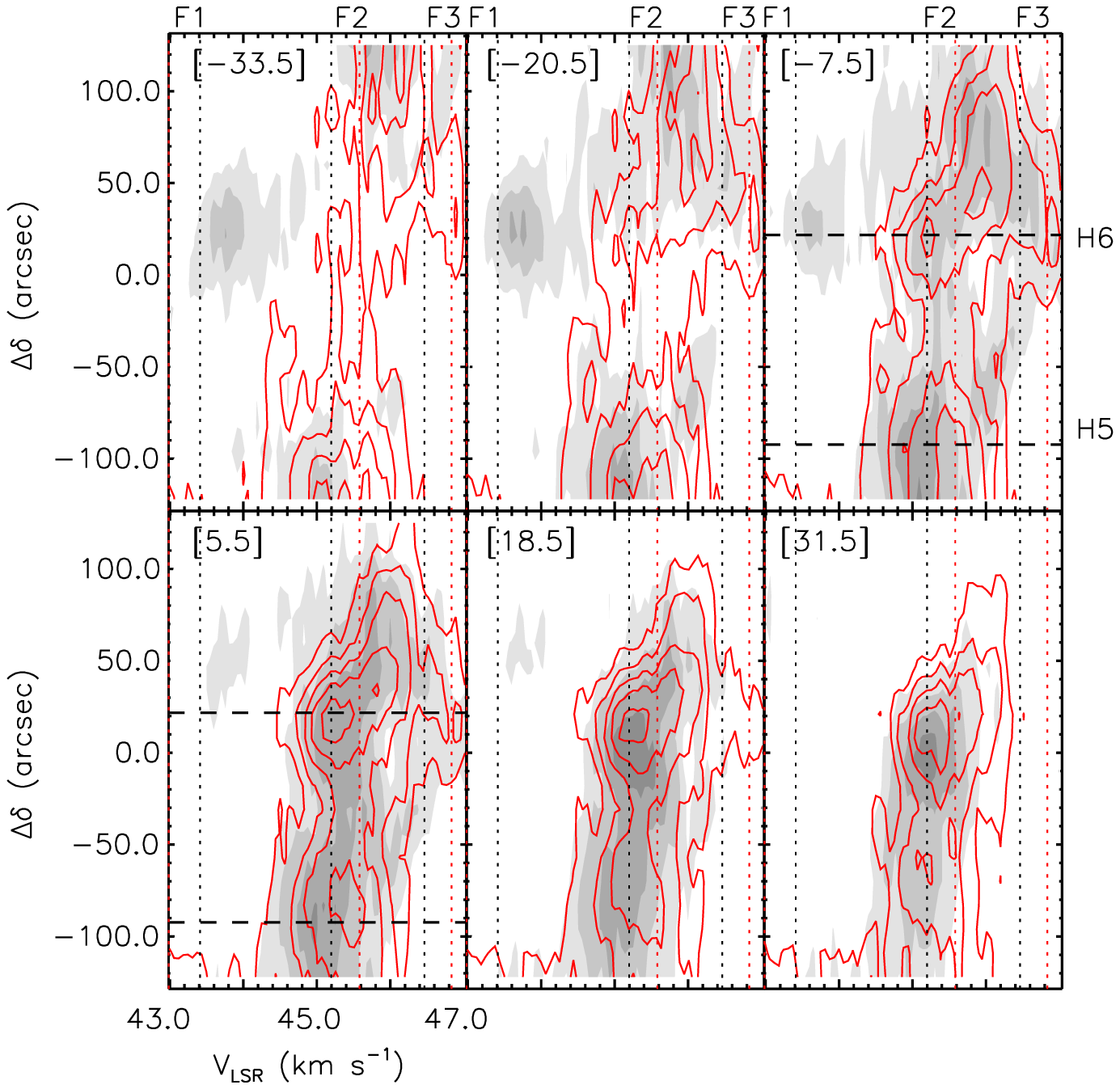}
\end{center}
\caption{Position-Velocity diagrams of both (Red) \ntwoh \ (\tonenj)
  (F$_1$, F = 0,1 $\rightarrow$ 1,2) isolated component and
  (Grey-scale) \co \ (\tonenj). Cuts are taken from North to South
  through the cloud at each offset right ascension (which can be found
  in the top left of each panel). Emission in \ntwoh \ (\tonenj) and
  \co \ (\tonenj) is plotted from 3$\sigma$ ($\sigma$ = 0.14, and for
  \ntwoh \ (\tonenj), $\sigma$ = 0.06). Contours increase in 3$\sigma$
  steps. Vertical dotted lines refer to the centroid velocities
  derived from the GGF (see Appendix \ref{GGF}). Horizontal dashed
  lines at offsets $\Delta\alpha$ = -7.5 and 5.5, refer to the
  positions of core H6 (2, 20), and core H5 (6, -86).}
\label{PV_contour_NS}
\end{figure*} 
%figure 7

Figure\,\ref{spectra_map} displays the individual \ntwoh \ (\tonenj)
(F$_1$, F = 0,1 $\rightarrow$ 1,2) and \co \ (\tonenj) spectra over
the full extent of the region mapped in \ntwoh \ (\tonenj). This is
overlaid on top of the mass surface density map, smoothed to
26\arcsec, to match the angular resolution. As previously noted in
Paper I, multiple velocity components appear at various positions
within the cloud (e.g. around core H6 and toward the South). This is
more apparent in \co \ (\tonenj) as it traces the more abundant, less
dense gas along the line of sight. However, the line morphologies of
the \co \ (\tonenj) are echoed in the South and Central positions in
\ntwoh \ (\tonenj). The fact that the profiles of the \ntwoh
\ (\tonenj) line are similar to the optically thin \co \ (\tonenj),
means that we are indeed seeing multiple velocity components along the
line of sight and not optical depth effects. In particular, {\it the
  {\rm \ntwoh \ (\tonenj)} spectra around the peak of core H6 (see
  offsets (-7.5, 8), (-7.5, 20), (5.5, 8), and (5.5, 21)), with a
  stronger blue peak typically observed in collapsing low-mass dense
  cores (e.g. \citealp{crapsi_2005}), are just the superposition of
  different velocity components.}

\begin{figure*}
\begin{center}
\includegraphics[trim = 40mm 35mm 65mm 40mm, clip
  ,width=1.0\textwidth]{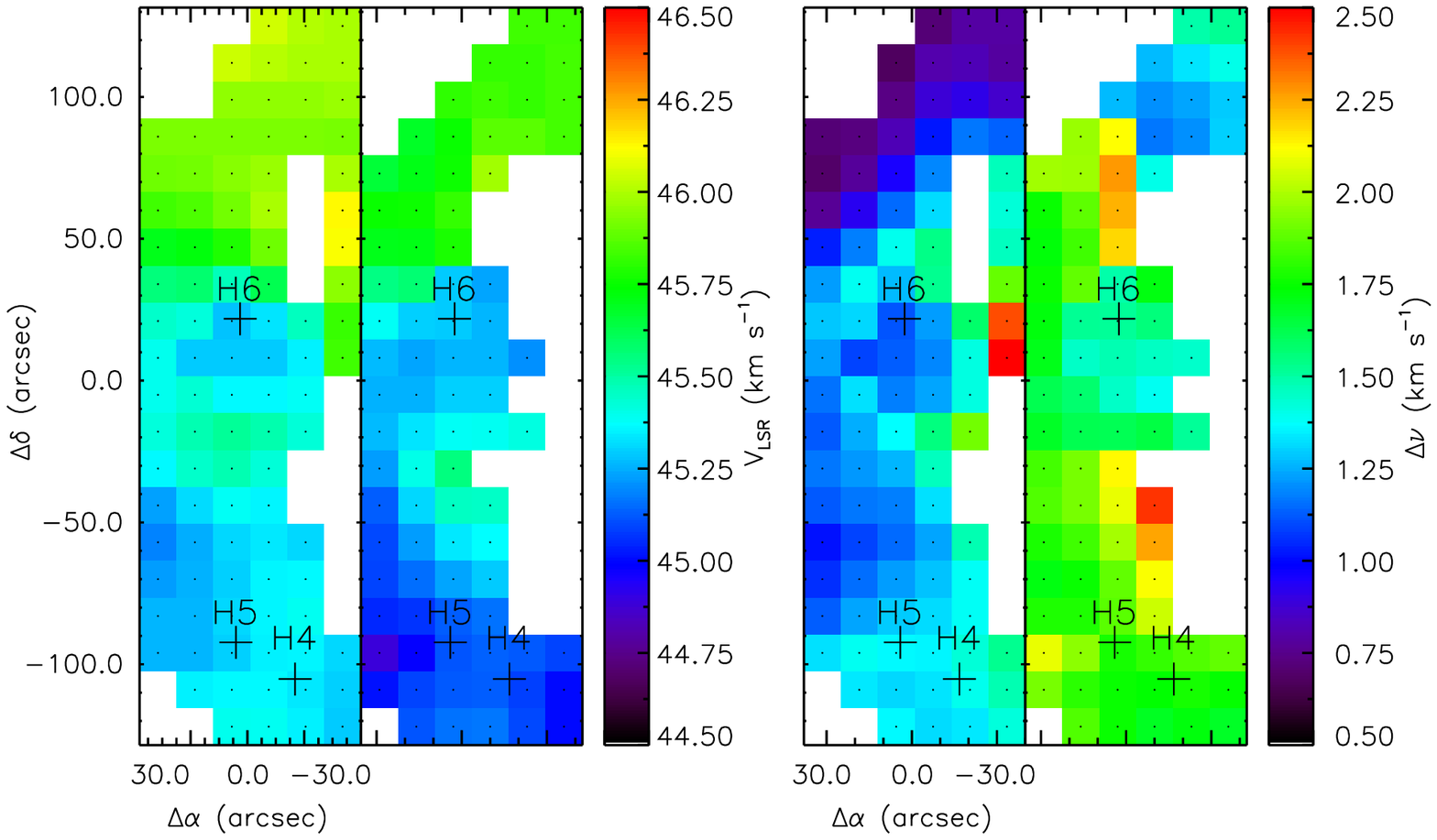}
\end{center}
\caption{V$_{LSR}$ and $\Delta\upsilon$ maps of Filament 2, as derived
  from the Guided Gaussian Fits (see Appendix \ref{GGF}). Maps on the
  left represent the V$_{LSR}$ of (Left) \ntwoh \ (\tonenj) and
  (Right) \co \ (\tonenj). Maps on the right represent the linewidth,
  $\Delta\upsilon$ of (Left) \ntwoh \ (\tonenj) and (Right) \co
  \ (\tonenj). Black crosses indicate the positions of the massive
  cores from BT12.}
\label{Vlsr_DV_maps}
\end{figure*}
%figure 8

\begin{figure*}
\begin{center}
\includegraphics[trim = 0mm 0mm 5mm 20mm, clip
  ,width=0.8\textwidth]{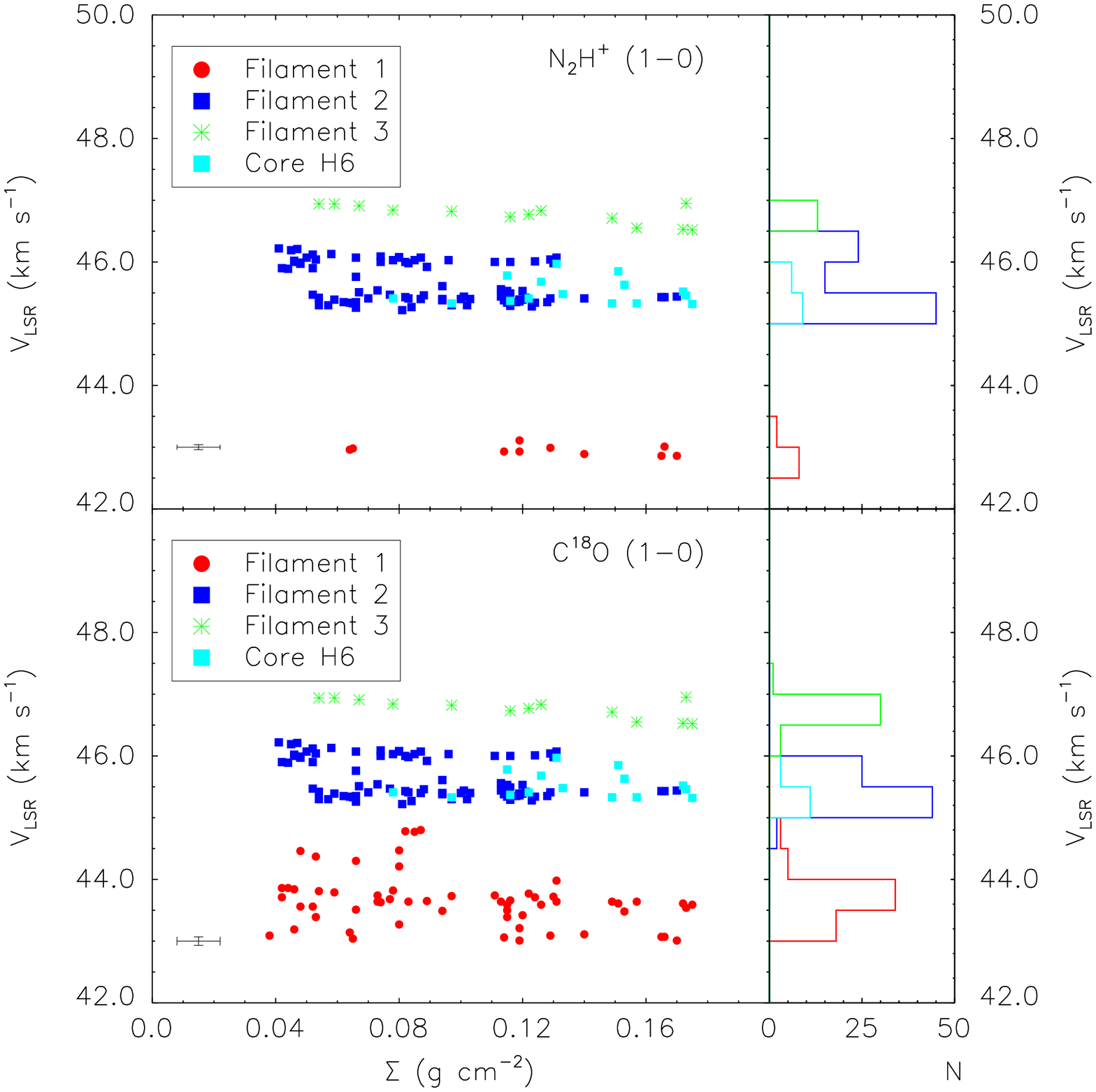}
\end{center}
\caption{V$_{\rm LSR}$ as a function of mass surface density for (Top)
  \ntwoh \ (\tonenj) and (Bottom) \co \ (\tonenj). The right of each
  plot shows a histogram of the V$_{\rm LSR}$ information. The
  uncertainty in the V$_{\rm LSR}$ are derived from the Gaussian fits
  and their mean values overall the three filaments are $\simeq$
  0.07\,\kms \ and 0.04\,\kms \ for \co \ and \ntwoh,
  respectively. Errors in the mass surface density represent the
  1$\sigma$ level = 0.007 g\,cm$^{-2}$. Mean uncertainties are
  dipicted in the bottom left-hand corner of each plot.}
\label{Vlsr_ext}
\end{figure*}
%figure 9

A more detailed view of the gas distribution can be found in
Figure\,\ref {channel_maps}. Here we present the channel maps of the
emission seen in \ntwoh \ (\tonenj) (F$_1$, F = 0,1 $\rightarrow$ 1,2)
and \co \ (\tonenj), between 42-48\kms, in velocity increments of
0.2\kms, superimposed onto the mass surface density map smoothed to
26\arcsec. From the figure, it is evident that the additional
components seen in the spectra of Figure\,\ref {spectra_map}, have
varying morphologies (see also Paper I for a similar analysis using
\co \ (\ttwonj) data). The first of these components (Filament 1) can
be seen between 42-44\kms. In \co \ (\tonenj) the component is seen to
run from North-East to South-East, curving to the West toward the
centre of the map. The \ntwoh \ (\tonenj) emission in this component
is restricted to the most southern portion of the cloud. The gas
distribution changes shape between 44-46\kms \ (Filament 2), peaking
at the positions of the massive cores. Here, the \ntwoh \ (\tonenj)
follows the morphology of the \co \ (\tonenj) more closely, and is
extended over the whole filament, indicating the presence of
widespread dense gas. Both species follow the shape of the extinction
map, from North-West to the South, curving slightly to the
East. Finally, another component is seen between 46-48\kms \ (Filament
3). This emission is similar in morphology to the previous
component. However, the emission is restricted to the northern portion
of the cloud, and is not seen in the South. Given that this component
has a well defined feature in both the spectra (Figure\,\ref
{spectra_map}), and the channel maps (Figure\,\ref {channel_maps}), we
consider this to be a separate filament.

Another way to inspect the gas kinematics across the cloud is through
position-velocity (PV) diagrams. Figure\,\ref {PV_contour_NS} shows
the PV diagrams of both species (\ntwoh \ (\tonenj) (F$_1$, F = 0,1
$\rightarrow$ 1,2) in red, and \co \ (\tonenj) in grey scale), slicing
the cloud from North to South, at each offset of right ascension in
our \ntwoh \ (\tonenj) map. Vertical dotted lines refer to the
velocities of the individual components seen in both species as
deduced by the Guided Gaussian Fit (GGF) of the average spectrum (see
Appendix A for further details). At the most westerly point of the
cloud ($\Delta\alpha$ = -33.5\arcsec), the \ntwoh \ (\tonenj) emission
follows the \co \ (\tonenj) emission in both the North and South,
except at $\Delta\delta$ $\sim$ 30\arcsec, where we can see an
additional component in \co \ (\tonenj) which is part of Filament
1. There is no significant velocity structure, except for a velocity
change of about 1\kms \ between the northern and southern emission on
a scale of $\sim$ 1 pc, which corresponds to a very small velocity
gradient of $\simeq$ 1\kms\,pc$^{-1}$.  Such a small gradient is
comparable to those found in low-mass cores
(e.g. \citealp{caselli_2002b}), but here it is seen over a
significantly larger extent.

As we move towards the centre of the cloud, the emission becomes
stronger (max intensity of \ntwoh \ (\tonenj) (F$_1$, F = 0,1
$\rightarrow$ 1,2) at -33.5\arcsec \ $=$ 0.82\,K\kms \ compared to
1.25\,K\kms \ at 5.5\arcsec) and spreads over a larger velocity
range. This can be seen towards offsets $\Delta\alpha$ = -20.5\arcsec
\ and -7.5\arcsec, where an additional component at higher velocities
(Filament 3) appears (most prominent between 0\arcsec \ $<$
$\Delta\delta$ $<$ 50\arcsec). Core H6 lies between -7.5\arcsec \ $<$
$\Delta\alpha$ $<$ 5.5\arcsec \ at $\Delta \delta$ = 20\arcsec. Here
all three filaments are seen, with Filament 2 being the most prominent
in \ntwoh \ (\tonenj). The channel maps of Figure\,\ref {channel_maps}
also show that these features coincide spatially at the position of
core H6, suggestive of filament merging (see below).

\begin{figure}
\begin{center}
\includegraphics[trim = 30mm 40mm 30mm 30mm, clip
  ,width=0.5\textwidth]{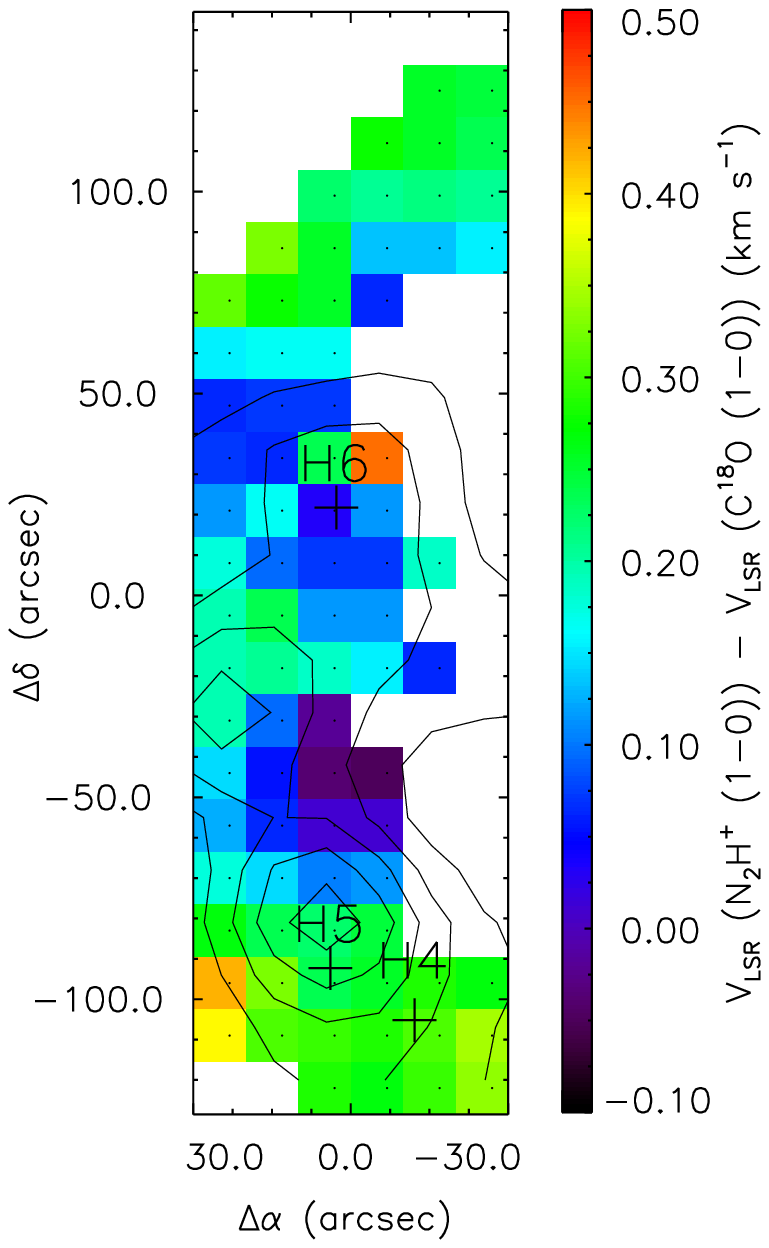}
\end{center}
\caption{Map of the velocity shift between \ntwoh \ (\tonenj) and \co
  \ (\tonenj) shown for Filament 2 only. Overlaid in black contours is
  the SiO (\ttwonj) emission. Contours are 3$\sigma$ to 0.5\,K\kms
  \ in steps of 3$\sigma$ ($\sim$ 0.1\,K\kms). Black crosses refer to
  the positions of the massive cores from BT12.}
\label{DVlsr_maps}
\end{figure}
%figure 10

\begin{figure}
\begin{center}
\includegraphics[trim = 5mm 10mm 5mm 5mm, clip
  ,width=0.5\textwidth]{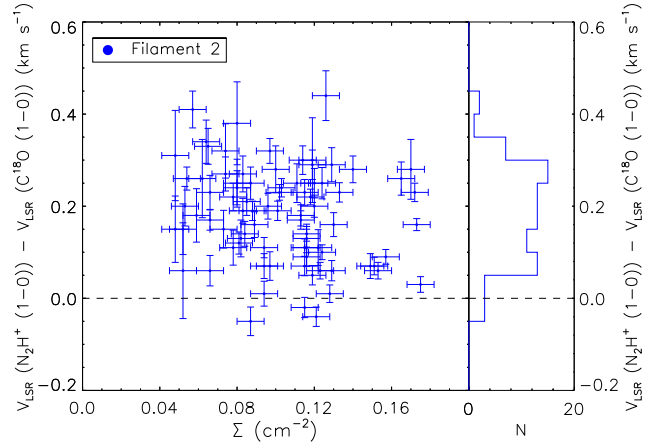}
\end{center}
\caption{Velocity shift between \ntwoh \ (\tonenj) and \co \ (\tonenj)
  as a function of mass surface density. The right of the plot shows a
  histogram of this velocity shift information. Errors in the mass
  surface density represent an assumed uncertainty of
  0.007\,g\,cm$^{-2}$. }
\label{DVlsr_ext}
\end{figure}
%figure 11

\subsubsection{Centroid velocity and line width}\label{GGF_results}

Figure\,\ref {Vlsr_DV_maps} shows the centroid velocity map of
Filament 2 in \ntwoh \ (\tonenj) (F$_1$, F = 0,1 $\rightarrow$ 1,2)
and \co \ (\tonenj)\footnote{Both the low and high velocity components
  (Filament 1 and 3, respectively) do not cover a large enough area in
  \ntwoh \ (\tonenj) to see any significant velocity gradient, so we
  do not consider the other two filaments in this analysis.}. These
velocities are derived from the GGF (Guided Gaussian Fit) method
(Appendix A). The velocity field appears coherent, except for a
velocity change of $\sim$ 0.5\kms \ at the northern edge of core
H6. The change in velocity from $\sim$ 45.9\kms \ (North) to $\sim$
45.4\kms \ (South) occurs within $\sim$ one beam width, corresponding
to a local velocity gradient of $\simeq$1\kms\,pc$^{-1}$. Same
conclusions are drawn from the \co \ (\tonenj) line, although the
velocity field appears more complex south of core H6. Thus, the abrupt
(but relatively small) velocity change is present in the dense
material as well as in the lower density envelope, suggesting that it
originates from the larger scale.

Figure\,\ref{Vlsr_DV_maps} also shows the \ntwoh \ (\tonenj) (F$_1$, F
= 0,1 $\rightarrow$ 1,2) and \co \ (\tonenj) line width maps of
Filament\,2. The first thing to note is that line widths are dominated
by non-thermal motions. In fact, adopting an average kinetic
temperature of 15\,K (e.g. \citealp{ragan_2011}), the thermal width
for both species is 0.15\kms. Thus, the observed widths are between 3
and 15 times the thermal width, with the narrowest widths found toward
the North.  On average, \co \ (\tonenj) and \ntwoh \ (\tonenj) line
widths have similar values.  {\it This is in striking contrast to
  low-mass star-forming regions, where lines of \co \ (\tonenj)
  typically show broader line widths compared to those measured in
  high density tracers (by a factor between 1.5 and 2;
  e.g. \citealp{fuller_1992})}. However, this is consistent with our
finding that \ntwoh \ is widespread across the filament and not
preferentially tracing dense cores as in low-mass star-forming regions
(see Figure 2). In both species, the line widths get narrower towards
the northern portion of the cloud and (to a lesser extent) toward the
starless core H6, indicating the presence of relatively quiescent (and
probably pristine) gas.

Figure\,\ref {Vlsr_ext} displays the V$_{LSR}$ of \ntwoh \ (\tonenj)
(F$_1$, F = 0,1 $\rightarrow$ 1,2) and \co \ (\tonenj) as a function
of mass surface density. The three filaments are clearly seen as well
separated velocity components (as determined by the Gaussian fits).
The average velocities of the individual components differ slightly
between the two species. In Filament 1, we find a mean V$_{LSR}$ of
42.95 (0.17)\kms \ in \ntwoh \ (\tonenj), and 43.65 (0.12)\kms \ in
\co \ (\tonenj). Filament 2 has a mean velocity of 45.63 (0.03)\kms
\ in \ntwoh \ (\tonenj), and 45.40 (0.03)\kms \ in \co
\ (\tonenj). Filament 3 has a mean velocity of 46.77 (0.06)\kms \ in
\ntwoh \ (\tonenj), and 46.76 (0.05)\kms \ in \co \ (\tonenj).  We
note that at mass surface densites below 0.06\,g\,cm$^{-2}$, the \co
\ filaments span a broader range of velocities. Moreover, the relative
velocity between Filament 2 and 1 is larger than that between Filament
2 and 3. This could be at the origin of the observed velocity shift
between \ntwoh \ and \co \ in Filament 2 (see
Figure\,\ref{spectra_map} and sections \ref{velocity_shift} and
\ref{discussion} for more details).  The abrupt change in velocity
observed in Filament 2 (Figure\,\ref{PV_contour_NS} and
\ref{Vlsr_DV_maps}) is seen here as the ``gap" between the two groups
of blue points.  It is interesting to see that {\it the velocities
  associated with the massive starless core H6 are filling this gap,
  suggesting that this core has probably formed at the interface of
  material moving at different velocities}.

\subsubsection{The {\rm \ntwoh - \co} velocity shift} \label{velocity_shift}

Figures\,\ref {DVlsr_maps} and \ref{DVlsr_ext} make clear that the
velocity shift seen between \ntwoh \ (\tonenj) and \co \ (\tonenj) is
the result of large-scale kinematics.  Figure\,\ref {DVlsr_maps} is a
map of the \ntwoh \ -- \co \ velocity shift measured across Filament
2, with the contours of SiO (\ttwonj) overlapped on top. The \ntwoh
\ (\tonenj) emission is largely red-shifted with respect to the \co
\ (\tonenj) emission, and we report a mean velocity shift of 0.18
$\pm$ 0.04\kms, across the whole filament. The largest velocity shifts
are seen to the North and South of the cloud. North of offset
$\Delta\delta$ = 86.0, the average velocity shift is 0.22
(0.04)\kms. South of offset $\Delta\delta$ = -70.0\arcsec, the
velocity shift is 0.26 (0.04)\kms. Between these offsets, the velocity
shift is 0.13 (0.04)\kms. No correlation is found between the velocity
shift and the SiO (\ttwonj) integrated intensity. The velocity shift
is also clear in Figure\,\ref {DVlsr_ext}, which shows the velocity
difference between \co \ and \ntwoh \ as a function of mass surface
density. Also in this case, no correlation is found, but {\it the
  shift is clearly present all across the filament}.

The velocity shift is not constant along the
filament. Figure\,\ref{DVlsr_fil} shows the \ntwoh \ -- \ \co
\ velocity difference as a function of declination along the four
central strips at fixed right ascension offsets -7.5\arcsec,
5.5\arcsec, 18.5\arcsec \ and 31.5\arcsec. From this figure it is
clear that the shifts in the North and South have similar magnitudes,
while variations are present in between. The variations along the four
strips appear similar up to the H6 core, where the outer strip at
$\Delta \alpha$ = -7.5\arcsec \ (and to a lesser extent the strip
passing through the peak of core H6) show different behavior. This is
the region where the abrupt change in velocity has been found (see
Figure\,\ref{Vlsr_DV_maps}). Those regions where the dense gas and its
envelope appear to be moving with the same velocity are present at
$\Delta \delta$ = -40\arcsec \ in the two western strips and at the
peak of core H6. However these points of minimum shift are surrounded
by sharp variations, especially evident toward core H6.  Discussion on
a possible origin of this velocity shift can be found in section
\ref{discussion}.

\subsection{The \co/\ntwoh \ integrated intensity ratio}

\citet{tafalla_2004} used CO depletion as a chemical clock, to
indicate an evolutionary sequence of individual cores. This is
achieved by taking two species that show different amounts of
depletion at high densities. \co \ freeze-out becomes heavy at
densities upwards of a few 10$^4$ cm$^{-3}$, whereas \ntwoh
\ (\tonenj) remains in the gas phase up to densities
$\geq$\,10$^5$\,cm$^{-3}$ (e.g.,
\citealp{caselli_1999,tafalla_2002}). Moreover, \ntwoh \ takes longer
than \co \ to form, due to the slower neutral-neutral reactions
involved in the production of N$_2$, the parent species of \ntwoh
\ (e.g. \citealp{hily-blant_2010}). Therefore, an early stage of
evolution would be characterised as having an undepleted \co \ and low
abundances of \ntwoh. As the cores follow their evolution toward star
formation, their density increases and \co \ starts to
freeze-out. Thus, with time, the CO abundance falls while \ntwoh
\ abundances reach equilibrium and maintain relatively large values up
to volume densities of $\ge$\,10$^5$\,cm$^{-3}$, as the freeze-out of
CO initially boosts the production rate of \ntwoh \ (given that CO is
an important destruction partner of \ntwoh).  In light of this, we can
use the method of \citet{tafalla_2004} and take the ratio of the
integrated intensities of \co \ (\tonenj) and \ntwoh \ (\tonenj):
\begin{equation}
R \equiv \bigg[\frac{I ({\rm C^{18}O})} {I ({\rm N_2H^{+}}) }\bigg],
%R = \bigg[\frac{C^{18}O}{N_2H^+}\bigg],
\end{equation}
to define a chemical clock. Chemically young gas is characterized by
values of $R$ $>$ 1. Figure \ref {int_ratio_map} displays the map of
the \co/\ntwoh \ intensity ratio integrated between 44 and
46\,km\,s$^{-1}$, where Filament 2 is more prominent. To obtain this
map, we first smoothed the \co \ (\tonenj) to 26\arcsec, the angular
resolution of the \ntwoh \ (\tonenj) map, and regridded the \co
\ (\tonenj) map. The \co/\ntwoh \ integrated intensity ratio has a
value $<$ 1 throughout the whole filament, except for the northern
regions.  The bulk of the filament shows large volume of \ntwoh \ with
respect to \co, and this suggests that the gas is in a chemically
evolved state.  This is clearly seen in Figure\,\ref{ratio}, where $R$
is plotted versus the mass surface density: no $R$ values larger than
1.0 are found at $\Sigma$ $\geq$ 0.09\,g\,cm$^{-2}$. Given that the
chemical time scale is inversely dependent on the number density, {\it
  the observed advanced chemical evolution is probably just the result
  of the relatively large average density of the filament when
  compared to low-mass star-forming regions} (see Section\,\ref{vol
  den}). Only in the North, the gas appears relatively less chemically
evolved, suggesting the presence of lower density material.

\subsection{Column Density}\label{col den}

\begin{figure*}
\begin{center}
\includegraphics[trim = 0mm 0mm 5mm 5mm, clip
  ,width=1.0\textwidth]{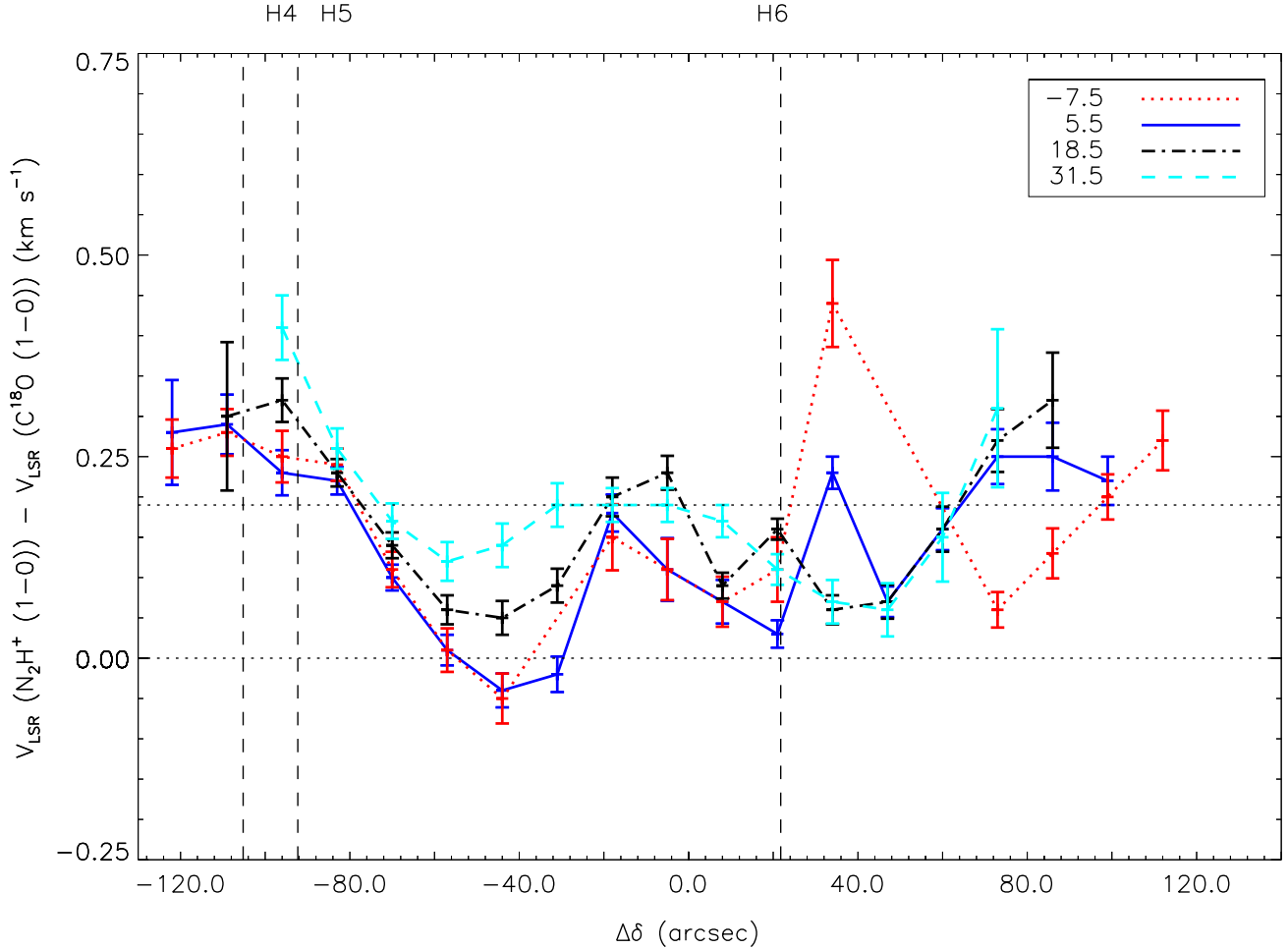}
\end{center}
\caption{Velocity shift between \ntwoh \ (\tonenj) and \co \ (\tonenj)
  as a function of offset in declination along four strips of constant
  right ascension (see legend in top right). The blue line refers to
  the strip passing closer to the center of core H6.  Horizontal
  dotted lines correspond to the mean values of the velocity shift,
  0.18$\pm$0.04\kms, and the zero line. Vertical dotted lines indicate
  the positions in offset declination of the three massive cores from
  BT12.  }
\label{DVlsr_fil}
\end{figure*}
%figure 12

The \ntwoh \ and \co \ column densities have been calculated from the
integrated intensity of the \ntwoh \ (\tonenj) F$_1$, F = 0,1
$\rightarrow$ 1,2 hyperfine component (divided by its statistical
weight) and \co \ (\tonenj) emission. The assumption here is that both
lines are optically thin. This has been verified for the \ntwoh
\ isolated component (with statistical weight of $\simeq$\,0.11111;
\citealp{caselli_1995}) with the HFS fitting technique (see Appendix
B) and for \co \ (\tonenj), Paper II). To measure the column
densities, we then used equation (A4) in Caselli et al. (2002).

The average \ntwoh \ column density across the IRDC is N(\ntwoh) \ =
(1.14 $\pm$ 0.11) $\times$ 10$^{13}$ cm$^{-2}$, peaking in the South
at offset (5.5\arcsec, -122\arcsec), with a value of (2.33 $\pm$ 0.25)
$\times$ 10$^{13}$ cm$^{-2}$. The minimum value of N(\ntwoh) is
located in the North at offset (5.5\arcsec, 125\arcsec), where
N(\ntwoh) \ = (2.33 $\pm$ 0.69) $\times$ 10$^{12}$ cm$^{-2}$. These
column densities are similar to those measured toward low-mass cores
\citep{caselli_2002b}, indicating that the average physical and
chemical conditions of the whole filament traced by \ntwoh (\tonenj)
are similar to those of the well-known dense cores in nearby low-mass
star-forming regions, in agreement with the study of
\citet{vasyunina_2011}.

The average \co \ column density throughout the filament is N(\co) =
(3.82 $\pm$ 0.04) $\times$ 10$^{15}$ cm$^{-2}$, peaking at position
(-10.0, 34.0), North-West of the extinction peak of core H6 with a
value of (5.95 $\pm$ 0.72) $\times$ 10$^{15}$ cm$^{-2}$. This is
consistent with Paper II, that found evidence of CO freeze-out toward
the center of H6.

We find a significant correlation (r-value = 0.8) between the column
density of \ntwoh \ (\tonenj) and the mass surface density (as
expected given the strong correlation found in
Figure\,\ref{ii_scatter} between the \ntwoh \ (\tonenj) integrated
intensity and the mass surface density): N(\ntwoh) \ =
(1.5$\pm$0.2)$\times$10$^{12}$ +
(9.8$\pm$0.3)$\times$10$^{13}$\,$\Sigma$. This correlation implies a
constant fractional abundance of \ntwoh \ (w.r.t. H$_2$ molecules) of
$\simeq$4$\times$10$^{-10}$, similar to the value found in low-mass
cores (a few times 10$^{-10}$; \citealp{caselli_2002a,crapsi_2005})
and in the quiescent massive starless core in Orion (Orion B9;
\citealp{miettinen_2012}), but lower than the value measured in the
direction of massive IR-dark clumps (a few $\times$10$^{-9}$;
\citealp{miettinen_2011}). Quiescent dense core conditions once again
appear to be appropriate in describing \irdc.

\subsection{Number density and kinetic temperature}\label{vol den}

Observations of the \tone \ and \tthree \ rotational transitions of
\ntwoh \ can constrain the number density, if the gas temperature is
known (see Section\,\ref{spectra}). No gas temperature measurements
have been made yet across \irdc, so we assume a constant temperature
of 15\,K across the cloud, based on the kinetic temperature derived in
other IRDCs (e.g. \citealp{pillai_2006,ragan_2011}), and close to the
dust temperature measured across the filament by \cite{nguyen_2011}
using Herschel data.

To calculate the number density, the non-LTE radiative transfer code
RADEX \citep{vandertak_2007} was used.  The \ntwoh \ (\tthreenj) map
was smoothed to 26\arcsec \ resolution, to match that of the \ntwoh
\ (\tonenj) data. Given the complexity of the spectra and the fact
that the (\tthreenj) transition does not have isolated hyperfine
components, we fitted a single Gaussian peak to the isolated hyperfine
component of the \ntwoh \ (\tonenj) irrespective of the fact that we
may see multiple components along the line of sight. In addition to
this, we performed a HFS fit assuming optically thin conditions to the
\ntwoh \ (\tthreenj) lines in CLASS, which gives, as one of its
outputs, the linewidth. The average linewidth between the \ntwoh
\ (\tonenj) and (\tthreenj) profiles was chosen as input parameter for
the RADEX calculation. The column density input to RADEX was
calculated using the integrated intensity of \ntwoh \ (\tonenj) over
the velocity range 40-50\kms \ (first using the isolated component and
then scaling by the statistical weight). As a final assumption we then
used a value of 15\,K for the kinetic temperature of the gas. The
number density was then varied between 1$\times$10$^3$ cm$^{-3}$ and
1$\times$10$^8$ cm$^{-3}$ and a minimisation technique was used to
find the best fit to the observed brightness temperature ratio of the
\tthree \ and \tone \ lines. Figure\,\ref{density} shows the result of
this minimization technique, i.e. the number density map of \irdc. The
average H$_2$ number density across the filament (measured with a
26\arcsec \ beam, corresponding to a linear size of 0.36\,pc) is
$\simeq$5$\times$10$^4$\,cm$^{-3}$, with peaks close to the three
massive cores and East of core H6. This value is roughly a factor of
10 larger than that found using extinction mapping in Paper
III. Possible causes of this difference are unresolved clumping and
the fact that the analysis done with the \ntwoh \ data only includes
those points where the 3-2 line is observed, therefore it is biased
toward higher densities (as the 3-2 line has a critical density of
3$\times$10$^6$\,cm$^{-3}$).

To check possible temperature variations across the filament, we fix
the number density to the average value found with the previous method
and find the kinetic temperature map. Figure\,\ref{temperature} shows
the kinetic temperature map obtained from the \ntwoh \ (\tthreenj
)/(\tonenj) intensity ratio and the RADEX code, assuming a constant
number density across the filament of 5$\times$10$^4$\,cm$^{-3}$ (the
average value found in Figure \ref{density}). A temperature gradient
is present perpendicular to the filament, with the largest values
($\sim$40\,K), close to the central SiO peak (see Figure
\ref{DVlsr_maps}).  Although this analysis should be taken with
caution, given the simplistic assumption of constant density, we
speculate that the observed temperature increase may be linked to the
filament merging and the consequent shock which has produced the
extended SiO emission (Paper I). In fact, the extinction in these
regions is still large enough to exclude any temperature enhancement
due to the interstellar radiation field. Independent temperature
measurements (e.g. from the inversion transitions of ammonia) and
higher angular resolution observations are needed to put more
stringent constraints on the temperature and density structure across
\irdc.

\begin{figure}
\begin{center}
\includegraphics[trim = 15mm 40mm 35mm 30mm, clip
  ,width=0.5\textwidth]{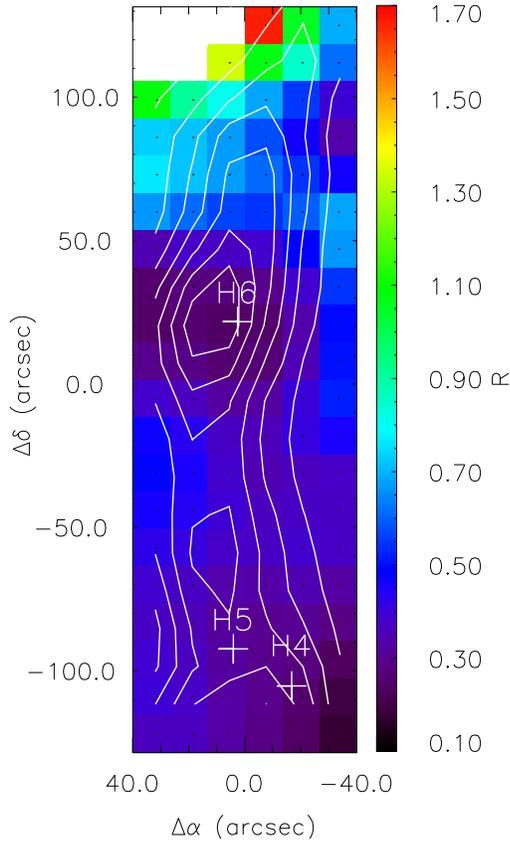}
\end{center}
\caption{Map of the integrated intensity ratio, R $\equiv [{I ({\rm
        C^{18}O})}/ {I ({\rm N_2H^{+}})} ]$. Intensities integrated
  between 44 and 46 \kms. Here, the integrated intensity of the
  isolated component of \ntwoh \ (\tonenj) is scaled up by its
  statistical weight, such that it is representative of the total
  integrated intensity. Overlaid on top are contours of the mass
  surface density smoothed to a 26\arcsec \ beam. Contour levels are
  3$\sigma$ ($\sigma$ $\sim$ 0.021 gcm$^{-2}$), increasing in steps of
  3$\sigma$. Crosses indicate the positions of the three massive cores
  are overlaid from BT12.}
\label{int_ratio_map}
\end{figure}
%figure 13

\begin{figure}
\begin{center}
\includegraphics[trim = 10mm 0mm 0mm 20mm, clip
  ,width=0.5\textwidth]{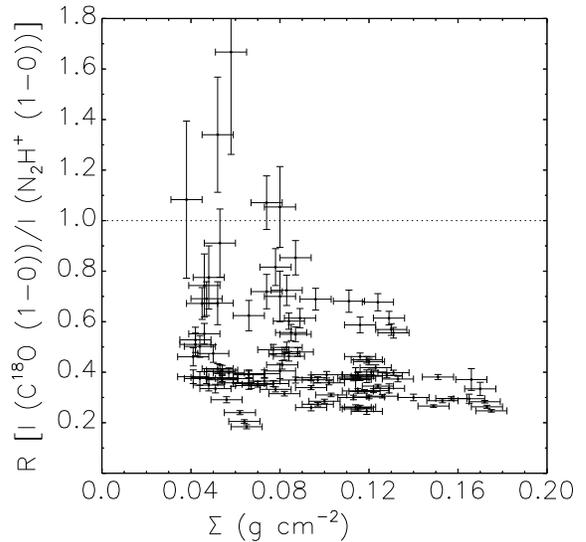}
\end{center}
\caption{Integrated intensity ratio, R $\equiv [{I ({\rm C^{18}O})}/
    {I ({\rm N_2H^{+}}) }]$, as a function of mass surface
  density. Values or $R$ less than 1 indicates chemically evolved
  material (see text). The observed $R$ drop is consistent with CO
  being depleted at high densities due to the freeze-out.}
\label{ratio}
\end{figure}
%figure 14

\begin{figure}
\begin{center}
\includegraphics[trim = 35mm 35mm 25mm 30mm, clip
  ,width=0.5\textwidth]{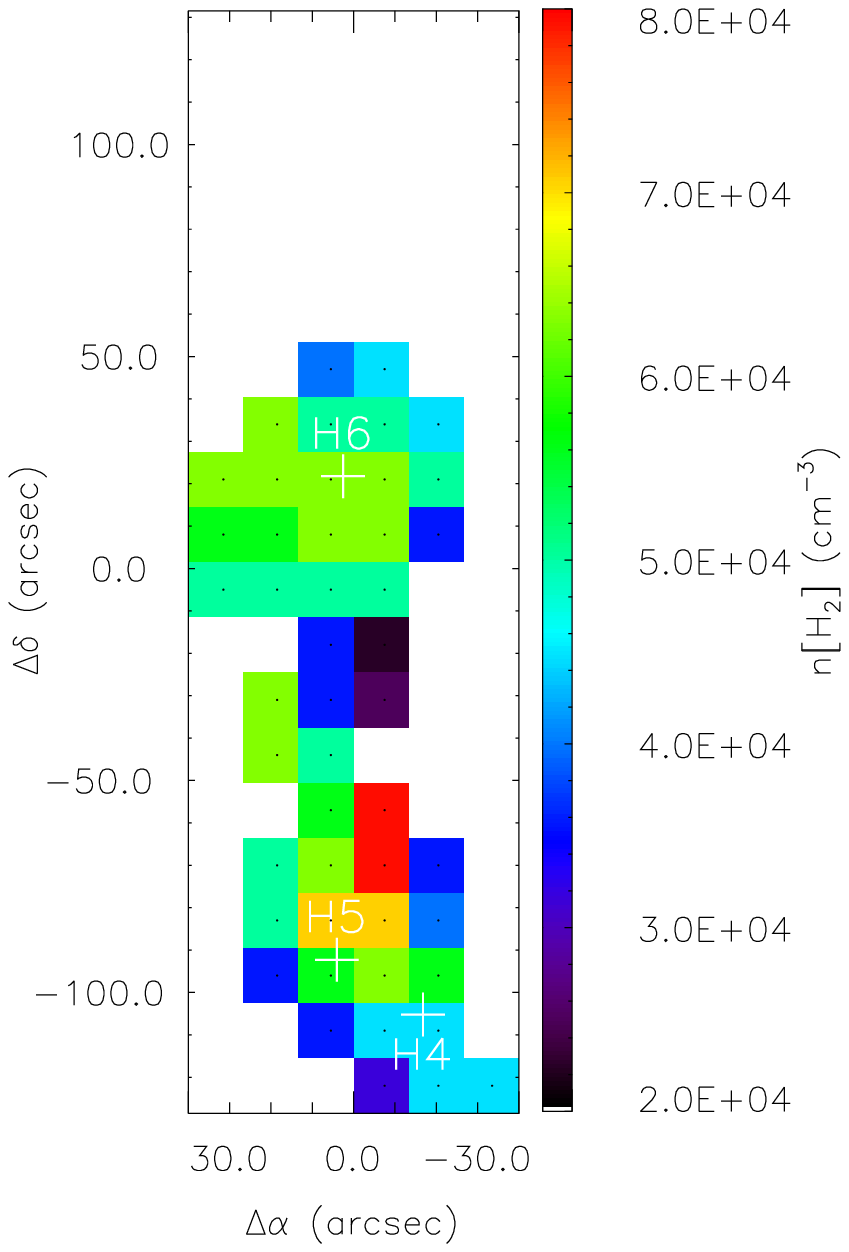}
\caption{Number density map of \irdc \ obtained from the observed
  \ntwoh \ (\tthreenj)/(\tonenj) line intensity ratio and the use of
  the RADEX code. A constant kinetic temperature of 15\,K is assumed
  across the cloud. Some regions at the edge of the filament show
  significantly higher density values than the average value,
  suggesting that the assumption of constant temperature may not be
  valid (see Figure\,\ref{temperature}). Crosses indicate the
  position of massive cores from BT12.}
\label{density}
\end{center}
\end{figure}
%figure 15

\begin{figure}
\begin{center}
\includegraphics[trim = 35mm 35mm 25mm 30mm, clip
  ,width=0.5\textwidth]{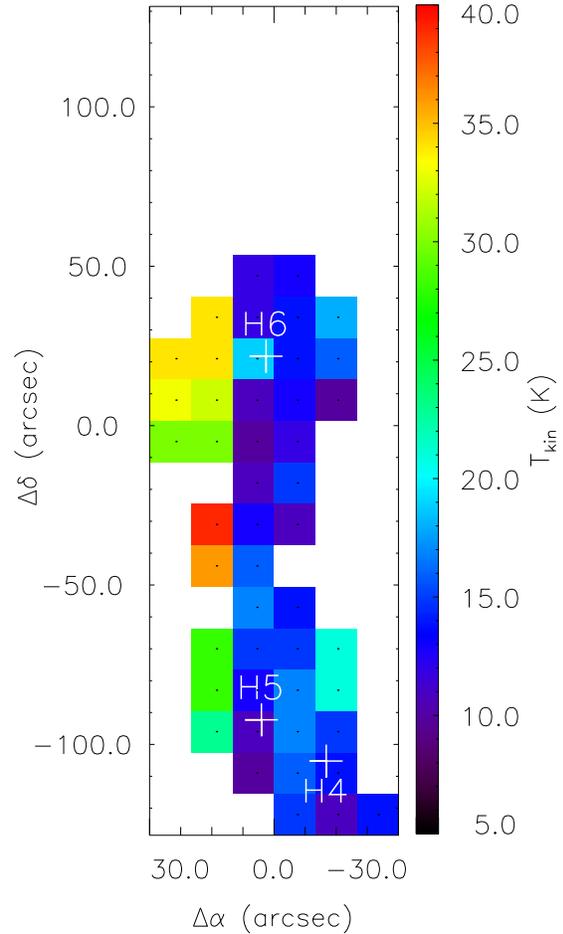}
\caption{Kinetic temperature map of \irdc \ obtained from the observed
  \ntwoh \ (\tthreenj )/(\tonenj) line intensity ratio and the use of
  the RADEX code. A constant number density of
  5$\times$10$^{4}$\,cm$^{-3}$ (the average number density in
  Figure\,\ref{density}) is assumed across the cloud. The line
  intensity ratio at the edge of the dense filament (Filament 2) is
  consistent with gas temperatures close to 20\,K. Crosses
  indicate the position of massive cores from BT12.}
\label{temperature}
\end{center}
\end{figure}
%figure 16

\section{Discussion}\label{discussion}

\subsection{What is causing the \ntwoh \ - \co \ velocity shift?}

The velocity shift between the \ntwoh \ (\tonenj), and \co \ (\tonenj)
reveals relative motions between the dense gas (traced by \ntwoh) and
the surrounding less dense envelope (traced by \co). This shift has
been searched for but not found in low-mass star-forming regions
(e.g. \citealp{walsh_2004,kirk_2007,hacar_2011}). \citet{hacar_2011}
studied the centroid velocities of \ntwoh \ (\tonenj) and \co
\ (\tonenj) in L1517, and found a good match between the two
tracers. They use this to say that there are no significant motions
between the different density regimes of the cloud, namely the dense
cores as traced by \ntwoh \ (\tonenj) and the surrounding envelope
traced by \co \ (\tonenj). They go on to conclude that this implies
that the velocity structure of the core is therefore not an intrinsic
property of the core, but a result of large-scale motions on
filamentary scales. In the case of \irdc, \ntwoh \ (\tonenj) traces
the dense cores as well as the filament over a distance of $\sim$ 4
pc, similar to the \co \ (\tonenj) line, so that the observed velocity
shift represents a puzzle which needs to be investigated in more
detail.

The channel maps in Figure\,\ref{channel_maps} clearly show that
significant \ntwoh \ (\tonenj) emission is present in a narrower range
of velocities compared to the \co \ (\tonenj) emission. In particular,
at the lowest velocities identified with Filament 1 (between 42 and
44\kms), \ntwoh \ is practically undetected along most of the
filament, whereas Filament 3 (between about 46 and 48\kms) is also
seen in \ntwoh \ (\tonenj) up to 47\kms.  If the three identified
filaments are interacting, the velocity field within the denser
Filament 2 will then be affected by Filament 1 and 3 in the case of
\co \ (\tonenj), but only by Filament 3 in the case of \ntwoh
\ (\tonenj).  This would imply a blueshift of the \co \ line, as we
indeed observe (see Figure\,\ref{DVlsr_ext}). The filament interaction is also evident in
Figure\,\ref{Vlsr_ext}, where the \co \ panel shows less distinct
velocity components, and Filament 2 is split in two velocity
components parallel to Filaments 1 and 3, with the densest region in
the IRDC (core H6) in between. The abrupt change in velocity seen at
the edge of core H6 (Figure\,\ref{Vlsr_DV_maps}) could also be
produced by filament merging, with the consequent formation of a dense
core at the intersection.  Finally, Figure\,\ref{DVlsr_fil} displays a
variation of the observed velocity shift between the \co \ (\tonenj)
and \ntwoh \ (\tonenj) lines along Filament 2, which can be explained
by inhomogeneities along the interacting Filaments 1 and 3. Indeed,
looking at Figures\,\ref{channel_maps}, \ref{DVlsr_maps} and
\ref{DVlsr_fil} one can see that the maxima in the velocity shift seen
in Filament 2 correspond to local maxima of the \co \ (\tonenj)
intensity in Filament 1. At these positions the densities are expected
to be higher (further indicated by the presence of \ntwoh \ (\tonenj))
and thus the interaction more prominent.

\begin{figure}
\begin{center}
\includegraphics[trim = 0mm 0mm 1mm 1mm, clip
  ,width=0.5\textwidth]{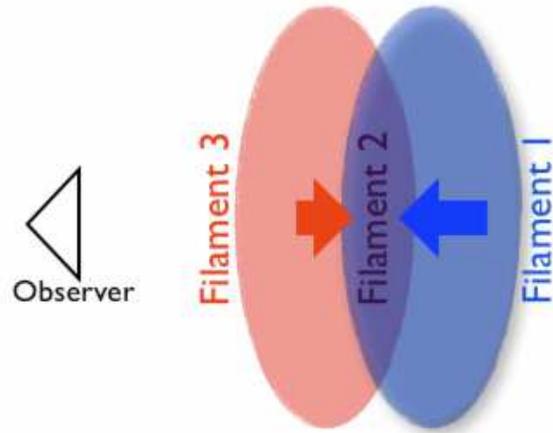}
\end{center}
\caption{Schematic figure illustrating the kinematics of \irdc. }
\label{sketch}
\end{figure}

The above observational evidence suggests that Filament 2 {\it may} be
the result of the merging of Filaments 1 and 3, as sketched in
Figure\,\ref{sketch}. This process of filament merging is still
ongoing, with material from Filament 1 still moving toward the denser
Filament 2 and causing the observed blueshift between the \co \ and
\ntwoh \ lines.  The relative velocity of Filaments 1 and 3 along the
line of sight is 3\kms \ (Figure\,\ref{Vlsr_ext}). Assuming that the
velocity components along the plane of the sky have similar
magnitudes, this implies a collision velocity of $\sim$ 5\kms.  The
(slow) shock which followed the impact has caused a density increase
in the interaction region. The (post-shock) density enhancement has
then contributed to the relatively large average density observed
across \irdc \ (Figure\,\ref{density}).  Before the collision, the
filaments appear to be quiescent and coherent in velocity, as
suggested by the relatively narrow lines and small velocity variations
observed in the North (Figure\,\ref{Vlsr_DV_maps}), where Filament 3
appears spatially separated from Filament 1. It is interesting to note
that the velocity coherence is maintained after collision, as
indicated by the presence of velocity gradients similar to those
measured in nearby low-mass dense cores ($\simeq$1\kms\,pc$^{-1}$, see
Section\,\ref{GGF_results}).

%The velocity shift observed between \ntwoh \ (\tthreenj ) and \ntwoh \
%(\tonenj) revealed by the average spectrum
%(Figure\,\ref{spectra_plot}) and that between \ntwoh \ (\tonenj) and
%\co \ (\tonenj) is consistent with a scenario of colliding filaments
%with different masses, where the densest gas (residing at the
%intersection) is found closer in velocity to the least massive
%filament (in our case Filament 3) instead of being at intermediate
%velocities, as expected in case of collision between similar mass
%filaments.

Note that, in this scenario, shock velocities of about 5\kms \ in gas
of $\simeq$10$^3$\,cm$^{-3}$ appear sufficient to inject enough
silicon in the gas phase to allow SiO to form and be detected across
\irdc \ (Paper I). However, these values of shock velocities and
densities are too low for sputtering and grain--grain collisions in
magnetized shocks to be effective in releasing silicon into the gas
phase (e.g. \citealp{caselli_1997, izaskun_2008, gusdorf_2008,
  guillet_2011}). It has been discussed in Paper I, that in order to
reproduce the observed fractional abundance of SiO via sputtering
within C-type shocks, some SiO has to be present in the icy mantles
(see also Schilke et al. 1997 and Jimenez-Serra et al. 2008), which
need a shock velocity slightly above 10\,\kms \ to be
sputtered. Therefore if sputtering is indeed the mechanism producing
the observed SiO emission, the \textit{total} velocity in the plane of
the sky would have to be at least 10\,\kms.  Similarly, in case of
grain-grain collisions within C-type shocks, the shock velocity has to
be at least 10\,\kms \ to allow dust grains with different electrical
charges to achieve a relative velocity larger than the threshold
velocity for icy mantle vaporization (6.5\,\kms , Tielens et al. 1994;
Caselli et al. 1997).

Alternatively, the release of SiO molecules mixed in icy mantles could
be the result of grain mantle vaporization due to grain--grain
collisions during the process of filament merging, without invoking
C-type shocks.  In this scenario, the grain-grain collision velocity
can be assumed close to the relative velocities of the merging
filaments, which is very close to the threshold velocity mentioned
above, especially taking into account the uncertainties in both the
observed relative velocity and the theoretical threshold velocity
value.  In fact, on the one hand, the inferred relative velocity of
$\sim$5\,\kms \ has been obtained assuming that the velocity
components along the plane of the sky have the same magnitude as the
component along the line of sight. On the other hand, the 6.5\,\kms
\ value given in Tielens et al. (1994) refers to water ice, with a
binding energy larger than 5000\,K.  If icy mantles are instead made
out of a mixture of H$_2$O and CO (as expected in regions with a gas
phase dominated by CO molecules), binding energy may be significantly
lower than 5000\,K and the vaporization threshold velocity will be
consequently lower. A factor of 5 decrease in binding energy
(appropriate for CO molecules bound on an water ice mantle; \"Oberg et
al. 2005) results in about a factor of 3 decrease in the threshold
velocity (using the formalism in Tielens et al. 1994). Thus, dust
grains moving at relative velocities of about 5\,\kms \ may lose a
significant fraction of their icy mantles upon collision.  A way to
test this scenario is to observe other species expected to be abundant
in the mantles of CO-rich dust grains, in particular CH$_3$OH, which
we expect to be widespread across the cloud.

\subsection{The Dynamical Evolution of \irdc}\label{evo}

The high-sensitivity and high-spectral resolution maps in both \ntwoh
\ (\tonenj) and \co \ (\tonenj) have revealed that the dense gas of
the \irdc \ is surrounded by complex, lower density filamentary
structures moving at relative velocities of a few \kms. Multiple
velocity components have also been detected in other IRDCs. For
instance \citet{devine_2011} found evidence for multiple components
separated by similar velocities in IRDC G19.30+0.07. In the previous
sections we have shown that in \irdc, these multiple filaments are
interacting.

Given this complexity of the IRDC, the kinematics are relatively
quiescent.  Toward the northern portion of the IRDC, where star
formation activity is at a minimum, the line widths are only a few
times the thermal width for a 15 K gas.  Here, \co \ (\tonenj) line
widths become close to 0.5\kms, similar to those measured in low mass
star-forming regions (e.g. \citealp{fuller_1992}).  In the rest of the
filament, the lines become a factor of 2-4 broader, probably due to a
combination of filament merging and embedded (but not yet prominent)
star formation activity (see also
\citealp{devine_2011,ragan_2012}). Moreover, the velocity structure
appears coherent, except toward core H6, where a 1\kms\,pc$^{-1}$
velocity gradient is measured (probably linked to the adjacent zone of
interaction).  Figure\,\ref {PV_contour_NS} shows no large velocity
gradients throughout the cloud. Hence, the merging of the filaments
must be fairly ``gentle'', possibly regulated by large-scale magnetic
fields. The presence of the velocity shift between \co \ and \ntwoh
\ lines is telling us that the accumulation of material toward the
region of interaction of the filaments is still ongoing. Assuming a
simplified 1D geometry, the time to build up the mass of Filament 2
from gas flowing in from Filaments 1 and 3 at relative velocity
$v_{\rm flow}$ is:

\begin{eqnarray}
t_{\rm flow} \sim 2\frac{R_f}{v_{\rm flow}} \frac{n_{\rm H, f}}
{n_{\rm H, flow}} \rightarrow \\ \nonumber{2.0
\bigg(\frac{R_f}{0.5\,{\rm pc}} \bigg) \bigg(\frac{v_{\rm
flow}}{5{\rm \,km\,s^{-1}}}\bigg)^{-1}\bigg(\frac{n_{\rm H,f}/n_{\rm
H,flow}}{10}\bigg)\,{\rm Myr}},
\end{eqnarray}
\label{tflow}

\noindent
where $R_f$ is the radius of the filament, here normalized to 0.5\,pc,
i.e. 36$^{\prime\prime}$ (see e.g. left panel of Fig.\ref{int_tot}),
$n_{{\rm H},f}$ is the mean total hydrogen nuclei number density in
Filament 2 and $n_{\rm H,flow}$ is the mean density in the merging
regions of Filaments 1 and 3. The ratio of these mean densities is
difficult to estimate, especially given the possibility of unresolved
clumping of the emission of particular gas tracers. From the
extinction map, Paper III estimated a density contrast of a factor of
3 for the case of the ``Inner Filament'' and its immediate envelope,
which is likely to be a lower limit to the ratio used in
Equation\,\ref{tflow}. Jimenez-Serra et al. (in prep.) find a density
contrast of a factor of $\sim$10, based on CO excitation analysis.
The timescale of $\ga$ 1\,Myr that is implied for the build-up of
Filament 2 by the converging flows of Filaments 1 and 3 is longer than
the local dynamical time in Filament 2 ($\simeq 0.8$\,Myr), a
necessary requirement if it has reached virial equilibrium as
concluded in Paper III.

%figure 17

We have proposed that the merging of filaments has led to the local
density increase along \irdc \ and, in particular at the position of
core H6. Cluster formation due to the collision of filaments has been
suggested in other star-forming regions, for instance in the massive
star-forming region W33A \citep{galvan_madrid_2010}, in the low-mass
star-forming region Serpens \citep{duarte_cabral_2011}, in the L1641-N
region \citep{nakamura_2012}, and in the Rosette Nebula
\citep{schneider_2012}. Here, we have been able, using high spectral,
high sensitivity data, to show that in this particular IRDC, the
merging of filaments appears to be occurring gently, at a relatively
low velocity, and that the build-up of material may go on for several
million years.

%In addition to cluster formation due to the collision and
%intersection of filaments, additional cores also form along the
%filaments. \citet{jackson_2010} have shown dense cores to be forming
%along the filamentary arms of the ``Nessie'' nebula, at a
%characteristic spacing of $\sim$ 4.5 pc, due to the ``varicose'' or
%``sausage'' instability. Dense core formation has also been
%discovered in addition to cluster formation along the filamentary
%structures in the Rosette nebula \citep{schneider_2012}. This mode of
%star formation is also not restricted to massive star-forming regions
%and has been seen in low-mass regions as well
%({e.g. \citealp{andre_2010,arz_2011}). Henshaw et al. (2012, in prep)
%will use higher resolution data from the Plateau de Bure
%Interferometer to test the paradigm that cluster formation is ongoing
%at the position of core H6, and further star formation is progressing
%along the filament.

\section{Summary \& Conclusions}\label{conclusions}

We have taken high sensitivity and high spectral resolution images of
the IRDC \irdc, chosen because of the cloud's characteristics
suggesting it is in an early stage of its evolution. From our analysis
we have found the following: \\

i) Rather than just the single prominent filament that is seen in
extinction, we find two additional, morphologically-distinct,
filaments with velocity components separated by a few separated by a
few \kms. In Filament 2, we find that \ntwoh \ (\tonenj) traces the
whole filament and not just the cores. There is a local maximum at the
position of core H6, and this massive core is spatially coincident
with the point of intersection of the velocity components. \textit{We
  suggest that the merging of Filaments 1 and 3 have produced the IRDC
  (Filament 2).}.

ii) Our kinematic study has revealed large-scale velocity coherence
(comparable line widths at most positions, lack of large velocity
gradients etc.) across the whole cloud. This is indicative of a
``gentle'' merging of filaments. \textit{We suggest that the merging
  of the filaments in the ongoing cloud formation process may be
  responsible for the large-scale SiO {\rm (\ttwonj)} emission seen in
  the \citet{izaskun_2010} study.}

iii) We have also witnessed for the first time, a widespread velocity
shift ($\sim$ 0.2\kms) between the \ntwoh \ (\tonenj) and \co
\ (\tonenj). Conversely, in low-mass regions, no velocity shift is
seen between these two tracers.  \textit{The presence of a velocity
  shift between these tracers over large scales is consistent with
  ongoing merging of filaments, which started $\ga$ 1\,Myr ago.} Star
formation within this cloud may have been locally induced by the IRDC
formation process (filament merging / cloud-could collision).

iv) Analysis using radiative transfer code, RADEX, gives an average
H$_2$ number density across the IRDC of about
5$\times$10$^4$\,cm$^{-3}$ and there is indication of a temperature
gradient perpendicular to the filament.

v) We find similarities and differences between the studied IRDC and
nearby (more quiescent) star-forming regions. Among the similarities
we list: (1) material that exhibits quiescent kinematics,
i.e. coherent velocities, especially in the northern regions where
star formation activity is at a minimum (suggesting that the filaments
have initially physical properties similar to those found in low-mass
star-forming regions); (2) small velocity gradients
($\simeq$1\kms\,pc$^{-1}$); (3) fractional abundances with respect to
H$_2$ molecules of \ntwoh \ $\simeq$4$\times$10$^{-10}$. Among the
differences we found: (1) non-thermally dominated motions traced by
\ntwoh, perhaps due to the fact that the dense gas in the IRDC has
been and continues to be formed from the collision of supersonically
interacting filaments; (2) large-scale emission of \ntwoh \ (\tonenj)
and (\tthreenj), implying relatively large densities ($n_{\rm H} \sim$
10$^5$\,cm$^{-3}$) extend over large, parsec-scale regions in the
IRDC.

vi) This study has also shown the importance of high spectral
resolution of different gas tracers to unveil the kinematics and
correctly interpret line asymmetries.  Asymmetric blue-peaked profiles
found across \irdc \ are due to multiple velocity components along the
line of sight. We did not find any evidence of large-scale infall
motions.  Instead, we find evidence of ongoing accumulation of
material at the intersection of two merging of filaments. Higher
angular resolution observations are needed to isolate single centers
of accretion (i.e. the star-forming dense cores analogous to the
well-studied low-mass cores).

\section*{Acknowledgements}
We acknowledge the IRAM staff for the help provided during the
observations. We thank Michael Butler and Jouni Kainulainen for
providing the extinction maps for this work. We thank Jaime E. Pineda
for assistance with image processing. PC work on star formation is
supported by a STFC rolling grant.
   
\bibliographystyle{mn2e}
\bibliography{references}

\appendix

\section{The Guided Gaussian Fit}\label{GGF}

\begin{figure}
\begin{center}
\includegraphics[trim = 50mm 50mm 20mm 10mm, clip
  ,width=0.5\textwidth]{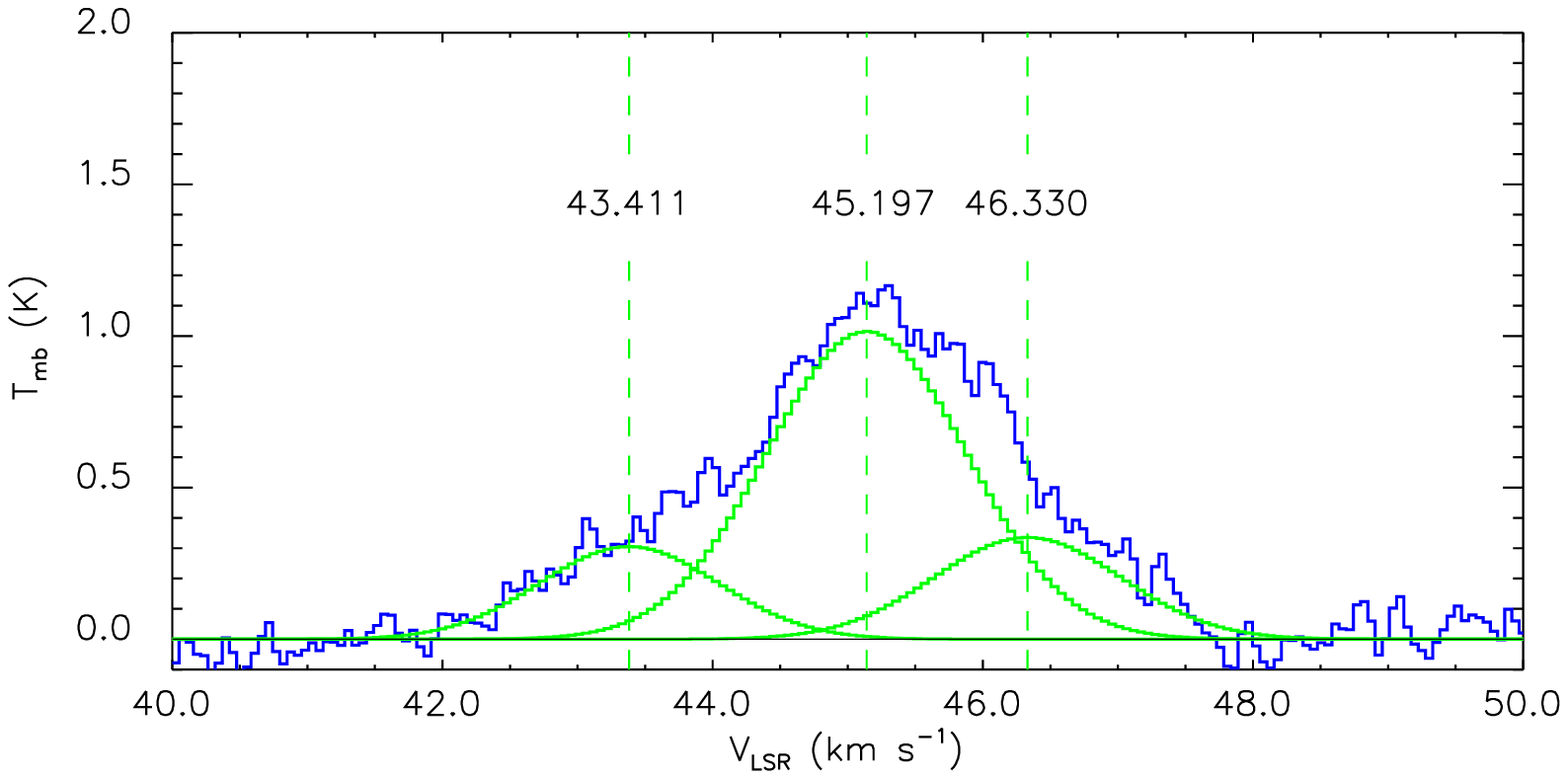}
\end{center}
\caption{Average spectra of the \co \ (\tonenj) shown in
  blue. Overlaid in green are the three components associated with the
  average spectrum at velocities, 43.411\,\kms, 45.197\,\kms, and
  46.330\,\kms.}
\label{A1} 
\end{figure}

Given the complexity of the data, we have developed a semi-automatic
procedure in order to fit and interpret the data, dubbed the Guided
Gaussian Fit (GGF). The fitting method is based on defining multiple
velocity components of gas along the line of sight. In the case of
\irdc, we observe three velocity components of differing morphologies
(see Section \ref {multiple vel} and Figures \ref {spectra_map} \&
\ref {channel_maps}). Based on the emission seen in Figure \ref
     {channel_maps} we define three windows in CLASS: 42-44 \kms,
     44-46 \kms, and 46-48 \kms. Using these three windows as a guide,
     we then fit three Gaussian profiles to the average spectrum of
     both species. The method here for the \co \ (\tonenj) is
     straightforward in that the three components are seen at numerous
     positions throughout the cloud and therefore are evident in the
     average spectrum. This result can be seen in Figure
     \ref{A1}. Here, the average spectrum of \co \ (\tonenj) is
     displayed with three Gaussian profiles overlaid. These peak at
     velocities of, (Filament 1) 43.411\,\kms, (Filament 2)
     45.197\,\kms, and (Filament 3) 46.330\,\kms. In the \ntwoh
     \ (\tonenj), Filaments 1 and 3 are only seen at a relatively
     small number of positions (compared to Filament 2). Thus, for the
     Filament 1 and 3 components of N$_2$H$^+$ we take the average
     spectrum over those positions where these components are
     detected, rather than over the full cloud, to avoid
     dilution. Centroid velocities, as determined by the GGF, for
     \ntwoh \ (\tonenj) of Filaments 1, 2, and 3 are 42.985\,\kms,
     45.582\,\kms, and 46.834\,\kms, respectively.

%Figure A1

By taking the line width output from the resulting fit, we can
calculate the velocity dispersion using,
\begin{equation}
{\sigma_V}_{LSR} = \frac{\Delta\upsilon}{2 \sqrt{2 \ln(2)}}.
\label{sig vlsr}
\end{equation}
We can now use this as the velocity window within which we search
for the three velocity components. CLASS is then used to calculate the
integrated intensity over these three windows. The standard detection
threshold for Gaussian profiles is the 3$\sigma$ level. In the case of
\irdc, we see three typically overlapping components along the line of
sight.  We have therefore held the detection limit for any individual
component at S/N $\geq$ 9.

\begin{figure*}
\begin{center}
\includegraphics[trim = 50mm 60mm 30mm 10mm, clip
  ,width=1.0\textwidth]{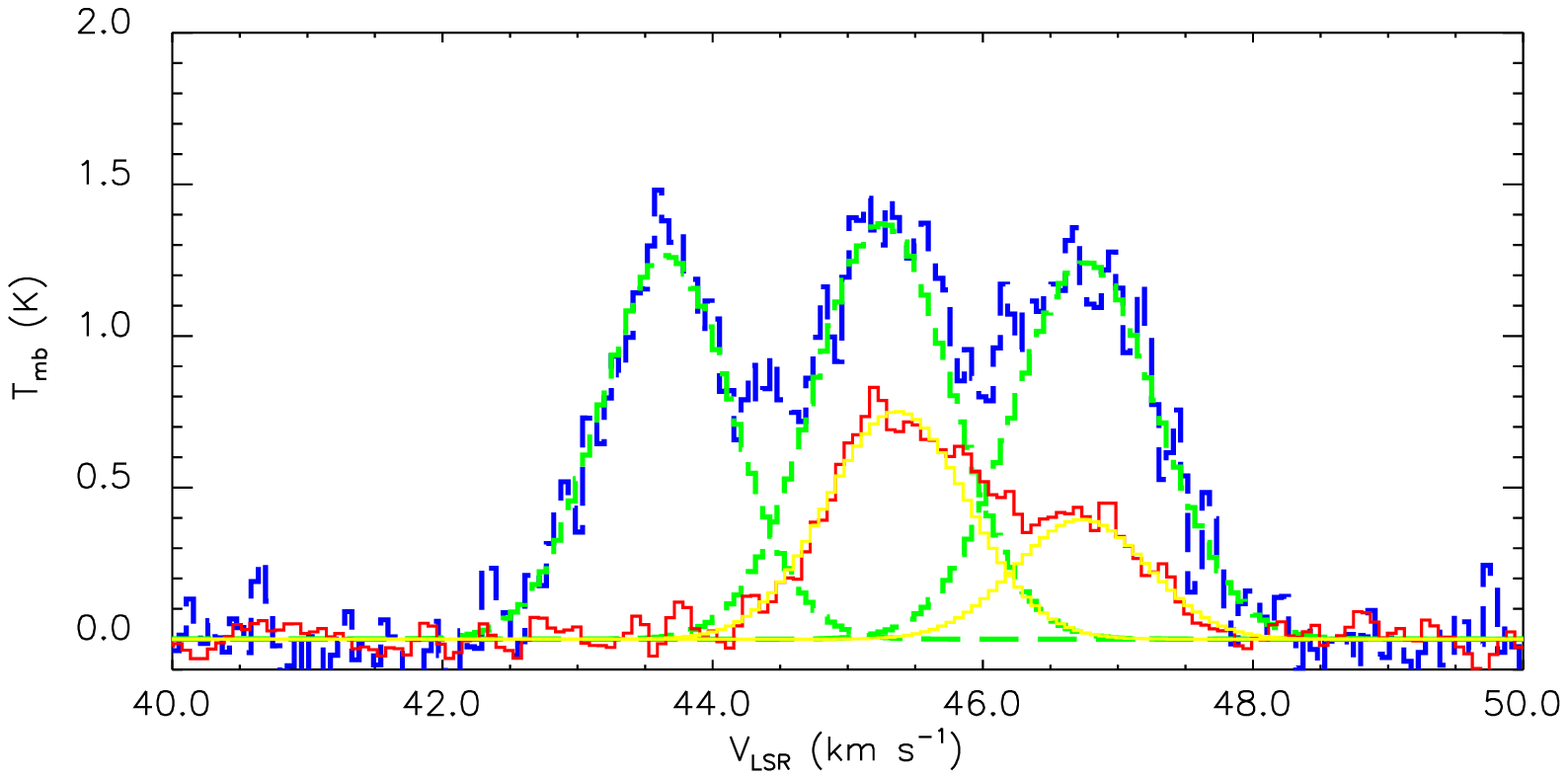}
\end{center}
\caption{Spectra at offset (-7.5, 21). \co \ (\tonenj) is shown as the
  dashed profile, in blue with the results of the GGF shown in
  green. \ntwoh \ (\tonenj) is shown as the solid profile, in red with
  the results of the GGF in yellow. }
\label{A2} 
\end{figure*}

Using this as a guide we now know for any given spectra (in \co
\ (\tonenj)), whether 1, 2 or 3 Gaussians should be fitted, and which
filaments should be fitted. In order to further verify this, the less
complicated isolated component of \ntwoh \ (\tonenj) is overlaid on
the spectra of the \co. Approximately, at any given position where
\ntwoh \ (\tonenj) is observed one can also expect to see \co
\ (\tonenj). Therefore, our ``extra-guide" to our kinematic study is
the \ntwoh \ (\tonenj) emission. If mutliple components are detected
in \ntwoh (\tonenj), and this is reflected in the spectra of the \co,
then the same components are recorded in both species. If a component
detected in \co \ (\tonenj) is not detected in \ntwoh \ (\tonenj),
then the \co \ (\tonenj) emission must be checked using the GGF
intensity and the gas distribution seen in Figure \ref
{channel_maps}. If indeed the feature meets these two conditions, it
is deduced to be a component in its own right and is fitted as
such. However, if the feature is indistinguishable as a component in
its own right, either due to blending of two filaments, or perhaps
depletion features, it is flagged as a blended line, and is removed
from the analysis. An example of this fitting proceedure can be seen
in Figure \ref{A2}, where by we plot the spectra of \co \ (\tonenj) in
blue, and \ntwoh \ (\tonenj) in red, at offset (-7.5, 21), at
approximately the position of core H6. The results of the guided
Gaussian fit are shown in green and yellow, respectively.

%Figure A2

\section{Gaussian and HFS fit results in selected positions}

\begin{table*}
\caption{Results of the guided Gaussian fitting method at selected
  offsets.\vspace{0.6cm}} \centering {\large
\begin{tabular}{c c c c c c}
\hline\hline
Position & Filament & T$_{mb}$ (K) & II (K \kms) & V$_{LSR}$ (\kms) & $\Delta\upsilon$ (\kms) \\ [0.5ex]
\hline
\multicolumn {6}{c}{\ntwoh (\tonenj)}\\
\hline
(-7.5, 21.0) & Filament 2 & 0.75 (0.04) & 0.98 (0.06) & 45.37 (0.03) & 1.23 (0.06) \\
... & Filament 3 & 0.40 (0.04) & 0.49 (0.05) & 46.73 (0.06) & 1.16 (0.11) \\
\\(5.5, 21.0) & Filament 2 & 0.93 (0.04) & 1.11 (0.03) & 45.32 (0.01) & 1.12 (0.02) \\
... & Filament 3 & 0.38 (0.04) & 0.61 (0.02) & 46.52 (0.02) & 1.52 (0.06)\\
\\(5.5, -31.0) & Filament 2 & 0.58 (0.04) & 0.78 (0.02) & 45.53 (0.01) & 1.26 (0.03) \\
\\(5.5, -83.0) & Filament 1 & 0.80 (0.03) & 1.10 (0.02) & 45.35 (0.01) & 1.29 (0.02)\\ [1ex]
\hline
\multicolumn {6}{c}{\co (\tonenj)}\\
\hline
(-7.5, 21.0) & Filament 1 & 1.27 (0.11) & 1.55 (0.07) & 43.66 (0.02) & 1.15 (0.05)\\
... & Filament 2 & 1.37 (0.11) & 1.75 (0.12) & 45.26 (0.02) & 1.20 (0.08) \\
... & Filament 3 & 1.24 (0.11) & 1.67 (0.09) & 46.76 (0.03) & 1.27 (0.07) \\
\\(5.5, 21.0) & Filament 1 & 0.78 (0.10) & 1.04 (0.05) & 43.59 (0.03) & 1.25 (0.07) \\
... & Filament 2 & 1.99 (0.10) & 2.37 (0.10) & 45.29 (0.02) & 1.12 (0.04) \\
... & Filament 3 & 1.13 (0.10) & 1.47 (0.08) & 46.72 (0.03) & 1.23 (0.07)\\
\\(5.5, -31.0) & Filament 2 & 1.65 (0.13) & 3.46 (0.06) & 45.55 (0.02) & 1.97 (0.04)\\
\\(5.5, -83.0) & Filament 2 & 2.01 (0.14) & 3.45 (0.06) & 45.13 (0.01) & 1.61 (0.04)\\ [1ex]
\hline\hline
\end{tabular}}
\label{obs table}
\end{table*}

\begin{table*}
\caption{Results of the HFS fitting method at selected offsets.\vspace{0.6cm}}
\centering
{\large
\begin{tabular}{c c c c c c}
\hline\hline
Position &  T$_{ant}$*Tau (K) & V$_{LSR}$ (\kms) & $\Delta\upsilon$ (\kms) & $\tau$ & T$_{ex}$ \\ [0.5ex]
\hline
\multicolumn {6}{c}{\ntwoh (\tonenj)}\\
\hline
(-7.5, 21.0) & 7.01 (0.13) & 45.74 (0.01) & 1.94 (0.02) & 3.48 (0.12) & 5.0 (0.1)  \\
(5.5, 21.0) & 9.06 (0.13)  & 45.63 (0.01) & 1.70 (0.01) & 3.42 (0.10) & 5.6 (0.1)  \\
(5.5, -31.0) & 5.14 (0.17) & 45.48 (0.01) & 1.40 (0.02) & 3.07 (0.23) & 4.6 (0.2)  \\ 
(5.5, -83.0) & 7.70 (0.15) & 45.38 (0.01) & 1.41 (0.01) & 4.28 (0.16) & 4.7 (0.1) \\ [1ex]

\hline
\multicolumn {6}{c}{\ntwoh (\tthreenj)}\\
\hline
(-4.0, 23.0) & 0.99 (0.06) & 45.83 (0.08) & 2.40 (0.19) & 0.10 & ... \\
(9.0, 23.0)  & 1.41 (0.10) & 45.43 (0.05) & 1.65 (0.16) & 0.10 & ... \\
(9.0, -29.0) & 0.61 (0.09) & 45.69 (0.09) & 0.81 (0.30) & 0.10 & ... \\ 
(9.0, -81.0) & 0.85 (0.07) & 45.36 (0.09) & 2.02 (0.20) & 0.10 & ... \\ [1ex]
\hline\hline
\end{tabular}}
\label{obs table2}
\end{table*}

\label{lastpage}
\end{document}